\documentclass[useAMS,usenatbib]{mn2e}
\usepackage{amsmath}
\usepackage{nccmath} 
\usepackage{stfloats}
\usepackage{epsfig}
\usepackage{natbib}
\usepackage{subfigure}
\usepackage{multirow}
\usepackage{amssymb}
\usepackage{gensymb} 
\def \Mdot 	{$ \text{M}_\odot $}
\def \Ldot 		{ \text{L}_\odot } 
\def \Ldotpc 	{\Ldot \ pc^{-2}}
\def \lbox 		{17cm}

\def \SFit  						{\log \ I_{\mathrm{e}} = -1.28\ \mathrm{log}\ R_{\mathrm{e}} + 1.32\ \mathrm{log}\ \sigma -0.29 }	
\def \MFit  						{\log \ I_{\mathrm{e}} = -1.31\ \mathrm{log}\ R_{\mathrm{e}} + 1.22\ \mathrm{log}\ \sigma -0.29 }				
\def \SFitPercent					{97\%}										
\def \MFitPercent					{93\%}					
					
\title{The Effect of Major Mergers on Age and Metallicity Across the Fundamental Plane}
\author[Porter et al.]{L.~A. Porter$^{1,2}\footnotemark[1]$, R.~S. Somerville$^{3}$, D.~J. Croton$^{4}$, M.~D. Covington$^{1,2,5}$, G.~J. Graves$^{6}$,\newauthor S.~M. Faber$^{7}$, and J.~R. Primack $^{1,2}$\\
     $^1$Department of Physics, University of California, Santa Cruz, California 95064, USA\\
	$^2$Santa Cruz Institute for Particle Physics, University of California, Santa Cruz, California 95064, USA\\	
     $^3$Department of Physics and Astronomy, Rutgers University, Piscataway, New Jersey 08854, USA\\
     $^4$Centre for Astrophysics and Supercomputing, Swinburne University of Technology, Melbourne, Australia\\
     	$^5$NSF International Research Fellow, Karst Research Institute ZRC SAZU, Titov trg 2, SI-6230 Postojna, Slovenia\\
	$^6$Miller Fellow, Department of Astronomy, University of California, Berkeley, CA 94720, USA\\
     $^7$ UCO/Lick Observatory, Department of Astronomy and Astrophysics, University of California, Santa Cruz, CA 95064, USA\\
     }

\begin{document}
\maketitle
\begin{abstract}
Recent low-redshift observations have attempted to determine the star formation histories of elliptical galaxies by tracking correlations between the stellar population parameters (age and metallicity) and the structural parameters that enter the fundamental plane (size $r_{e}$ and velocity dispersion $\sigma$).  These studies have found that velocity dispersion, rather than effective radius or dynamical mass, is the main predictor of a galaxy's stellar age and metallicity.  In this work, we apply an analytic model that predicts the structural properties of remnants formed in major mergers to progenitor disk galaxies with properties taken from two different semi-analytic models (SAMs).  We predict the effective radius, velocity dispersion, luminosity, age, and metallicity of the merger remnants, enabling us to compare directly to observations of early-type galaxies.  We note that our SAM population consists solely of elliptical galaxies whose evolution is artificially truncated following the formative major merger, while we are comparing to a low-redshift population of elliptical and S0 galaxies that may have more complex formation histories.  While we find a tight correlation between age and velocity dispersion, we find a stronger dependence of age and metallicity on effective radius than observations report.  We find that these correlations arise as a result of the dependence of gas fraction, age, and metallicity on the stellar mass in the progenitor disk galaxies.  These dependences induce a net rotation in the radius-velocity plane between the correlations with effective radius and circular velocity in the disk galaxy progenitors, and the correlations with effective radius and velocity dispersion in the elliptical galaxy remnants.  The differences between our results and those from observations suggest that major mergers alone are not sufficient to produce the observed lack of correlation between effective radius and stellar population parameters.  Simulations have suggested that subsequent minor mergers may introduce scatter in the effective radius while leaving the velocity dispersion essentially unchanged.  Incorporating such minor mergers into the model may, then, bring the simulations into closer agreement with observations.
\end{abstract}
\begin{keywords}
galaxies: interactions -- galaxies: evolution -- galaxies: elliptical and lenticular, cD -- galaxies: formation
\end{keywords}
	\footnotetext[1]{email: laporter@ucsc.edu}
\section{Introduction}

Simulations of galaxy formation have shown that elliptical galaxies may be created through major mergers of disk galaxies \citep[e.g.][]{Toomre:1972a,Toomre:1977a,Mihos:1994a}.  Small, compact ellipticals can form at high redshifts from gas-rich `wet' mergers \citep{Dekel:2006a,Ciotti:2007a}.  Subsequent dissipationless `dry' mergers greatly increase the galaxies' stellar masses and radii, producing the massive ellipticals seen in the local universe \citep{Khochfar:2003a, Naab:2006b,Naab:2009a,Oser:2012a}.

Early type galaxies can be described by a two-dimensional plane relating effective radius ($R_\mathrm{{e}}$), central stellar velocity dispersion ($\sigma$), and effective surface brightness ($I_{\mathrm{e}}$), termed the fundamental plane (FP) \citep{Djorgovski:1987b,Dressler:1987a,Faber:1987a}.  This plane is tilted from the plane one would expect from a simple application of the virial theorem, indicating that further processes, such as non-homology or a varying mass-to-light ratio, must have an effect \citep{Jorgensen:1996b}.  Furthermore, the fundamental plane is not an exact relation; galaxies have a degree of scatter around the FP, in effect making the fundamental plane `thick'. Observations indicate that this scatter increases with redshift, particularly among less massive galaxies \citep{Treu:2005b}.

More specifically, while the slope of the FP appears unchanged for high-mass ellipticals since $z \sim 1$, low-mass ellipticals at high redshifts have higher surface brightnesses than their effective radii and velocity dispersions would seem to predict \citep{Wel:2004a,Treu:2005a,Treu:2005b,Jorgensen:2006a,Dokkum:2007a}.  If we consider a projected FP, where surface brightness is the dependent parameter, then these low-mass galaxies tend to lie above the mean FP relation at high redshift, and to fall onto the FP over time.

There are indications that this residual thickness in the FP correlates with the stellar population age.  \cite{Forbes:1998a} and \cite{Terlevich:2002a} found that galaxies with higher residual surface brightnesses are younger than those that lie near the
mid-plane of the FP; conversely, those with lower residual surface brightnesses are older.

More recently, observations from the Sloan Digital Sky Survey (SDSS) have been used to analyze stellar population trends both within the R-$\sigma$ projection of the FP, and through the thickness of the FP, using residual surface brightnesses \citep{Graves:2009c,Graves:2009b,Graves:2010b,Graves:2010a}.  By stacking spectra of galaxies with similar stellar properties and measuring the Lick indices on those spectra, the authors were able to derive [Fe/H],[Mg/H], [Mg/Fe], and stellar age for a population of passive early-type galaxies. In agreement with \cite{Forbes:1998a} and \cite{Terlevich:2002a}, Graves et al.~(2009b, hereafter G09) found that younger galaxies lie above the FP, having relatively higher surface brightnesses, while older galaxies lie below it; galaxies above the FP also tended to have higher [Fe/H] and [Mg/H], and lower [Mg/Fe].  G09 also determined that age, [Fe/H], [Mg/H], and [Mg/Fe] increase with velocity dispersion throughout the FP, independent of the residual surface brightness.  These same properties are almost independent of $R_\mathrm{{e}}$, indicating that a galaxy's velocity dispersion is a more reliable determinant of its star formation history than its dynamical mass ($\propto \sigma^{2}R_{e}$).  The strong dependence of age and metallicity on velocity dispersion is consistent with previous studies \citep{Smith:2007a,Nelan:2005a}.  Similar results were obtained in a recent analysis of the Six-degree Field Galaxy Survey \citep{Jones:2004a,Jones:2009a} by \cite{Springob:2011a}, though this work finds a slightly higher dependence on effective radius.  G09 posited that the lack of dependence on $R_{e}$ occurs because effective radius is strongly dependent on the orbital parameters of the major merger and subsequent dissipationless mergers.  This dependence introduces a large amount of scatter in the remnants' radii, effectively diluting any corresponding radial relations that were present in the progenitors.

Semi-analytic models (SAMs) provide a framework to simulate the formation and evolution of elliptical galaxies in a cosmological context \citep{Kauffmann:1993a,Cole:1994a,Somerville:1999b,Cole:2000a,Hatton:2003a,Croton:2006a,De-Lucia:2006a,Bower:2006a,Somerville:2008a,Fontanot:2009a,Benson:2010a,Cook:2010a,Guo:2011a,Somerville:2011b}.  However, attempts to use SAMs to study the fundamental plane have been limited by the difficulty in modeling the effective radii and velocity dispersions of ellipticals.  \cite{Cole:2000a} provide a simple formula to predict the radii of elliptical galaxies following a major merger using the virial theorem and conservation of energy.  While this relation may be correct for dissipationless gas-poor mergers, the energy lost due to star formation in gas-rich mergers results in a deviation from the virial relation, and smaller remnant radii.  Incorporating this dissipation is probably essential: a recent study using the \cite{Bower:2006a} SAM framework could match the observed SDSS size-mass relation only by adding dissipation \citep{Shankar:2011a}. A second recent SAM paper that did not include dissipation in modeling merger remnants had a too shallow slope \citep{Guo:2011a}.  The predicted scatter in the size-mass relations in these papers was also much larger than observed.  Full N-body/SPH simulations of gas-rich mergers have been more successful \citep{Dekel:2006a,Robertson:2006a,Hopkins:2008a}, but they do not simulate enough galaxies to compare with observations of populations of galaxies as in G09.

Using hydrodynamical simulations \citep{Cox:2004b,Cox:2006a,Cox:2008b}, Covington et al.~(2008, hereafter C08) developed an analytic model to predict the effective radius and velocity dispersion following the major merger of two disk galaxies.  This model was based upon the virial theorem and additionally incorporated energy losses due to dissipation.  In further work, Covington et al.~(2011, hereafter C11) simplified the model and applied it to mergers of disk-dominated galaxies with properties taken from the \cite{Croton:2006a} Millennium SAM and the \cite{Somerville:2008a} SAM via post-processing.  C11 showed that the C08 model reproduces the observed steepening in the size-mass relation of ellipticals when compared to disks, as well as the observed decreased dispersion in radius compared to the progenitor galaxies \citep{Shen:2003a}.   In addition, this model qualitatively reproduced the evolution of these properties versus redshift \citep{Trujillo:2006a}. The C08 model was also shown to correctly reproduce a tilt in the FP away from the simple virial relation.  Using the methods of C08 and C11, as well as an alternative prescription from \cite{Hopkins:2009c}, \cite{Shankar:2011a} reached similar conclusions.

The predictions of effective radius and velocity dispersion from C08 and C11, combined with information from SAMs, allow us to determine a simulated galaxy's location above, within, or below the FP, as measured by its residual surface brightness.  In this work, we relate the age and metallicity of simulated galaxies across the FP with effective radius and velocity dispersion.  We compare galaxies with initial conditions taken from the \cite{Croton:2006a} and \cite{Somerville:2008a} SAMs to each other as well as to the analysis of G09, and find that the stellar population metallically is correlated with both radius and velocity dispersion, while age is strongly dependent on velocity dispersion and almost independent of radius.  The metallicity correlation is in contrast to observational results (G09), which found a tight correlation with velocity dispersion only and almost no correlation with radius for either age or metallicity.  We find that all five measured parameters (age, metallicity, luminosity, velocity dispersion and radius) are closely tied to those of the disk progenitors, with the main difference being a rotation in the two-dimensional $R_{\mathrm{e}}$-$\sigma$ projection of the fundamental plane due to differing amounts of dissipation in galaxies with different gas fractions.  We find no change in the FP-stellar population correlations with the redshift of the formative merger.  Examining the thickness of the fundamental plane, we find that galaxies that fall below the FP, with relatively lower surface brightnesses, tend to be both older and more metal-poor than galaxies that lie above the FP.  This is in agreement with observations \citep{Graves:2009b,Terlevich:2002a,Forbes:1998a}.  

Section \ref{ssec:SAMmodel} describes the two semi-analytic models used in our analysis.  Section \ref{ssec:Covmodel} provides an overview of the analytic model used to calculate the radius and velocity dispersion of the remnants of major mergers between two disk galaxies.  The prescription we use combines the more detailed model for effective radius and velocity dispersion from C08 with an improved model to predict the concentration of dark matter within the effective radius $R_{e}$ \citep{Covington:2011a}.  Sections 2.3 through 2.4 describe our calculations of the light-weighted ages, light-weighted metallicities, and luminosities of the remnants at redshift zero, using the mass-weighted ages and metallicities of the progenitors, as well as the stellar population models of \cite{Bruzual:2003b}.  We also describe the procedure for separating galaxies according to their location on the fundamental plane, based on their residual surface brightnesses.  Section 3 presents a summary of the low-redshift observations of G09, to which we make direct comparisons. 

We present results beginning in Section 4, in which we examine the relationships between age and metallicity, on the one hand, and velocity dispersion and radius, on the other, throughout the fundamental plane for the \cite{Croton:2006a} and \cite{Somerville:2008a} SAMs.  In both cases (age and metallicity), and in both SAMs, the relations for the remnants represent a net $\lesssim 20\degree$ counterclockwise rotation from relations for the progenitors; thus, correlations with radius (velocity dispersion) in the progenitors tend to produce mixed correlations involving both components in the remnants.  Section 5 examines the residual thickness of the FP and the structural differences between galaxies lying above and below the FP.  In Section 6 we compare our results to those of G09. Section 7 examines the evolution of the FP correlations as a function of the redshift of the formative merger event, and attempts to predict the effects of including subsequent evolution via multiple mergers, including minor mergers.    

\section{Methods}

We begin with an overview of the main components of the model: the Millennium \citep{Croton:2006a} and \cite{Somerville:2008a}, hereafter S08, semi-analytic models, from which we drew progenitor galaxies, and the dissipational merger model, which was used to predict the structural properties of the merger remnants.

\subsection {The semi-analytic models}

\label{ssec:SAMmodel}

We model elliptical galaxy formation using the properties of late-type progenitors, taken from SAMs, to predict the ages, metallicities, luminosities, velocity dispersions, and effective radii of elliptical remnants following a major merger.  Using the results of N-body dark matter simulations or the extended Press-Schechter formalism, the two SAMs construct merger trees of dark matter halos, which are then seeded with galaxies.  When two halos merge, their galaxies are considered to merge on a characteristic timescale, so long as the smaller galaxy is not tidally disrupted before the merger completes.  The evolution of this galactic substructure is treated in a semi-analytic manner, using simple prescriptions for photoionization, radiative cooling, star formation, supernova feedback, and black hole growth and feedback.  This methodology has been shown to accurately reproduce low-redshift observed relations, such as the stellar mass function, cold gas fractions, and specific star formation rates.  For more detail, we refer the reader to \cite{Somerville:2008a,Croton:2006a}.

Drawing from the Millennium and S08 merger trees, we select pairs of disk-dominated galaxies that merge with a stellar mass ratio equal to or greater than 1:3 and with a stellar mass greater than $10^9$  \Mdot.  We exclude mergers of bulge-dominated galaxies, since the SAMs do not currently produce accurate estimations of bulge size, and hence stellar radius, for these galaxies.  These mergers also tend to be dissipationless, as the bulge-dominated galaxy is usually both more massive and gas-poor; thus the baryonic gas fraction of the merger is relatively low.  Since we want to produce quiescent galaxies at redshift zero, we exclude any mergers from the past 1 Gyr ($ z < 0.075$).  Making these selection cuts, we include 57,435 mergers from the S08 SAM and 53,416 mergers from the Millennium SAM.  For simplicity, we do not consider any further evolution following the major merger; possible implications of subsequent evolution are discussed in section 7.1.  We also do not attempt to model the formation and growth of bulges via disk instabilities, or the formation of S0 galaxies.  In effect, we are only studying the relations elliptical galaxies would have if they formed in a major merger and then passively evolved to redshift zero.  

Both simulations use cosmological parameters taken from the first-year Wilkinson Microwave Anisotropy Probe (WMAP) results: $\Omega_m$ = 0.25, $\Omega_b$ = 0.045, $\emph{h}$ = 0.73, $\sigma_8$ = 0.90 \citep{Spergel:2003a}.

\subsection{Dissipational model for remnant structural parameters}
\label{ssec:Covmodel}
To locate elliptical remnants in FP-space, we have built upon the dissipational model of Covington et al. \citep{Covington:2008b,Covington:2011a}.  This is a simple analytic model to predict the effective radius and velocity dispersion following major mergers, derived from adding radiative energy loss to a standard dissipationless model \citep[see][]{Cole:2000a}.  Here we provide a brief summary of the derivations used to calculate the parameters we will be using, namely the remnant galaxy's stellar mass, half-mass radius, and velocity dispersion.  For a full description, we refer the reader to C08 and C11.

In a major merger, the efficiency of the burst of star formation depends on the amount of cold gas available and the dissipation due to the merger.  We parameterize the dissipational strength of each galaxy as $f_{k} \equiv \Delta E/K_{tot}$, where $\Delta E$ is the impulse between the two galaxies and $K_{tot}$ is the total initial kinetic energy of the galaxy. To determine the impulse, pericentric merger distances and velocities were drawn from the orbital parameter distributions given in \cite{Benson:2005a}, though the results do not change significantly if newer parameters \citep{Wetzel:2010a} are used.  The mass of new stars formed for each progenitor is given by:
\begin{fleqn}
	\begin{equation}	
		\label{eqn:Star_formation_mass}
		M_{\mathrm{new\ stars}}=C_{\mathrm{new}}M_{\mathrm{gas}}f_{\mathrm{k}}
	\end{equation}
\end{fleqn}
where $M_{\mathrm{gas}}$ is the mass of gas in the progenitor and $C_{\mathrm{new}} \sim 0.3$  is obtained by fitting to the \cite{Cox:2004b} simulations.

Equating the initial internal energies of the progenitors with the sum of the internal energy of the remnant and the radiated energy lost due to star formation, C08 found: 
\begin{fleqn}
	\begin{equation}
		\label{eqn:Energy_init}
		E_{\mathrm{init}}=C_{\mathrm{int}}G{\left(\frac{{(M_{\mathrm{s},1}+M_{\mathrm{ns},1})}^{2}}{R_{1}}+\frac{{(M_{\mathrm{s},2}+M_{\mathrm{ns},2})}^{2}}{R_{2}}\right),}
	\end{equation}
	\begin{equation}
		\label{eqn:Energy_conservation}
		E_{\mathrm{final}}=E_{\mathrm{init}}+E_{\mathrm{rad}},
	\end{equation}
	and
	\begin{equation}
		\label{eqn:Energy_final}
		E_{\mathrm{final}}=C_{\mathrm{int}}G{\left(\frac{{(M_{\mathrm{s},1}+M_{\mathrm{s},2}+M_{\mathrm{ns},1}+M_{\mathrm{ns},2})}^{2}}{R_{\mathrm{final}}}\right),}
	\end{equation} 
\end{fleqn}
where $M_{\mathrm{s},i}$ is the stellar mass of the progenitor, $M_{\mathrm{ns},i}$ is the mass of stars formed in the progenitor during the merger, $R_{i}$ is the stellar half-mass radius of the progenitor, and $C_{\mathrm{int}}=0.5$ is a constant tuned to the \cite{Cox:2004b} simulations.  

The radiated energy is given by:
\begin{fleqn}
\begin{equation}
		\label{eqn:Energy_radiated}
		E_{\mathrm{rad}}=\sum_{i=1}^{2}K_{i}f_{\mathrm{g},i}f_{\mathrm{k},i}(1+f_{\mathrm{k},i})
	\end{equation}
\end{fleqn}
where $K_{\emph{i}}$, $f_{\mathrm{g},i}$, and $f_{\mathrm{k},i}$ are the total kinetic energy, baryonic gas fraction, and fractional impulse of progenitor \emph{i}.  An additional term, corresponding to the energy of the orbit itself, is added to the conservation of energy calculation (equation \ref{eqn:Energy_conservation}); however this term is typically on the order of a few percent of the radiated and initial energy terms, and so does not significantly affect the remnant's radius. Using equations \ref{eqn:Energy_conservation} and \ref{eqn:Energy_final}, the half-mass radius of the remnant can be determined.

We use a modified virial-type relation to determine the velocity dispersion of the remnant:
\begin{fleqn}
	\begin{equation}
		\label{eqn:velocity_dispersion} 
		\sigma={\left(\frac{GC_{\mathrm{sig}}M_{\mathrm{s,f}}}{2R_{\mathrm{f}}(1-f_{\mathrm{dm,f}})}\right)}^{1/2},
	\end{equation} 
 \end{fleqn}
where $M_{\mathrm{s,f}}$ is the stellar mass of the remnant, $f_{\mathrm{dm,f}}$ is the central dark matter fraction of the remnant (i.e. the fraction of dark matter within the stellar half-mass radius), $R_{\mathrm{f}}$ is the half-mass radius of the remnant, and $C_{\mathrm{sig}}$ = 0.15 is a constant set to match the \cite{Cox:2004b} simulations.  The internal energy terms of equations \ref{eqn:Energy_init}, \ref{eqn:Energy_conservation} and \ref{eqn:Energy_final} use the total mass of the progenitors, rather than the baryonic mass.

\subsection {Calculating the light-weighted remnant age and metallicity}

As we interested in the effects of major mergers alone, we artificially freeze all the physical properties of the galaxy directly after the merger.  The stellar population is allowed to passively evolve, but we do not consider subsequent mergers, star formation, or mass accretion.  Both the S08 and Millennium SAMs provide mass-weighted ages and total metallicities of the progenitor galaxies at the time of the merger.  In order to directly compare to the G09 data, we convert the mass-weighted values to light-weighted ones, considering the remnant galaxy to be comprised of two progenitors and a new stellar population formed in a burst during the merger.  The weighting formula we have used is:
\begin{fleqn}
	\begin{equation}
		\label{eqn:Light_weighting}
		A_{\mathrm{L,R}} =\frac{ \Sigma(A_{\mathrm{M,P}}*L_{\mathrm{P}}) +A_{\mathrm{N}}*L_{\mathrm{N}}}{\Sigma(L_{\mathrm{P}})+L_{\mathrm{N}}}
	\end{equation} 
 \end{fleqn}
where $A_{\mathrm{L,R}}$ is the light-weighted age of the remnant, $A_{\mathrm{M,P}}$ is the mass-weighted age of the progenitor, $A_{\mathrm{N}}$ is the age of the new stars, and $L_\mathrm{P}$ and $L_\mathrm{N}$ are the V-band luminosities of the progenitor and new stars, respectively.  A similar formula is used to determine the light-weighted metallicity, where the metallicities of the progenitors and new stars are used in place of their ages.  The summations are over progenitor galaxies.  All ages and luminosities are calculated at a redshift of 0.0, to compare with G09.  
 
Since the stellar populations of the progenitor galaxies are assumed to passively evolve to redshift zero, we take their redshift zero mass-weighted ages to be equivalent to their light-weighted ages, setting the mass-to-light ratio to 1.  We assume all new stars form at the instant of the merger, so the age of the new stars at redshift zero is equivalent to the lookback time of the merger.
 
Following \cite{Croton:2006a} and \cite{Somerville:2008a}, we assume that stars formed during the merger have the same metallacity as the cold gas of their host galaxies at the onset of the merger.  While the semi-analytic models include prescriptions to model chemical evolution through supernova feedback and accretion from the inter-galactic medium, we do not include any further chemical evolution after the merger event.

\subsection{Computing the remnant luminosities}

The luminosities of the progenitors and remnants are determined using the stellar population synthesis models of \cite{Bruzual:2003b}, hereafter BC03.  As noted above, all calculations are done at a redshift of zero.  

We use a model based on the Padova 1994 stellar evolution tracks \citep{Bertelli:1994a} with a \cite{Chabrier:2003a} initial mass function. The progenitor galaxies are taken to be simple stellar populations (SSPs), with luminosities at the time of the merger provided from the respective SAMs.  We use these quantities to calculate the mass-to-light ratio, and hence the luminosity, at redshift zero using a linear interpolation of the BC03 models in log time and (roughly) log metallicity.  We exclude progenitors whose metallicities are outside the range of the BC03 models, 0.0001 $\le$ \emph{Z} $\le$ 0.05; these galaxies represent less than 0.3\% of the progenitor population and would not significantly affect the results.  

The calculation of the mass-to-light ratio for new stars formed during the merger is slightly more involved.  The star formation rate from the merger is modeled as exponentially decaying in time, and we use BC03 to form tracks based on the Padova 1994 evolution tracks for this population.  More specifically, the star formation rate is:
\begin{equation}\label{eqn:star_formation_rate}\psi(t) = [1 M_\odot - f_\mathrm{R} M_{\mathrm{PG}}(t)]\tau^{-1} \exp(-t/\tau) \end{equation}
where $M_{\mathrm{PG}}(t) $ is the mass of processed gas that has been converted into stars and returned to the interstellar medium, determined within BC03, and $f_\mathrm{R}$ is the fraction of $M_{\mathrm{PG}}$ recycled back into new star formation.  $\tau$ is the e-folding timescale; for the actual calculation we used an equivalent parameter, the efficiency of star formation: $e= 1- \mbox{exp}(-1\ \mbox{Gyr}/\tau)$.  Note that BC03 defines efficiency as the fraction of gas converted into stars after 1 Gyr, while the efficiency we use is the fraction of gas converted over the entire burst. We use $f_\mathrm{R} = 0.43$, which is appropriate for the adopted Chabrier IMF.

\subsection {Binning in the fundamental plane}

Once the age, metallicity, effective radius, luminosity, and velocity dispersion have been calculated for the remnants, we separate them
into five slices according to their location above or below the fundamental plane, using surface brightness as the independent variable.  Since we intend to compare to the G09 results, only galaxies that fall within the G09 range of radius, $\mathrm{(0.0 \ kpc < log\ \emph{R}_{e} <0.7\ kpc)}$, and velocity dispersion, $\mathrm{(2.0 \ km\ s^{-1} < log \ \sigma
  <2.4\ km\ s^{-1})}$, are included in the fitting routine.  We use a least-squares fit to determine a relation between (log) $R_{\mathrm{e}}$, (log) $\sigma$, and (log) $I_{\mathrm{e}}$, finding 
\begin{fleqn}
	\begin{equation}
		\label{eqn:S08_FP_params} 
		\SFit
	\end{equation} 
	for S08 and 
	\begin{equation}
		\label{eqn:Milli_FP_params} 
		\MFit
	\end{equation} 
\end{fleqn}
for Millennium. For each remnant, the predicted surface brightness is determined using the above relation, and galaxies are separated by
their residuals, where $\mathrm{(-0.3\ \Ldotpc <  \Delta\ log\ \emph{I}_{e} < -0.1\ \Ldotpc)}$,
$\mathrm{(-0.1\ \Ldotpc<\Delta\ log\ \emph{I}_{e} <0.1\ \Ldotpc)}$, and $\mathrm{0.1\ \Ldotpc<\Delta\ log\ \emph{I}_{e} <0.3\ \Ldotpc)}$
are termed the low-, mid-, and high-FP, respectively.  Galaxies with residuals outside the range $(\mathrm{-0.3\ \Ldotpc <
  \Delta\ log\ \emph{I}_{e} < 0.3\ \Ldotpc)}$ are placed in the very low (`vlow') and very high (`vhigh') FP slices.

\begin{figure*}
	\centering
		\subfigure{
			{\includegraphics[width=7cm]{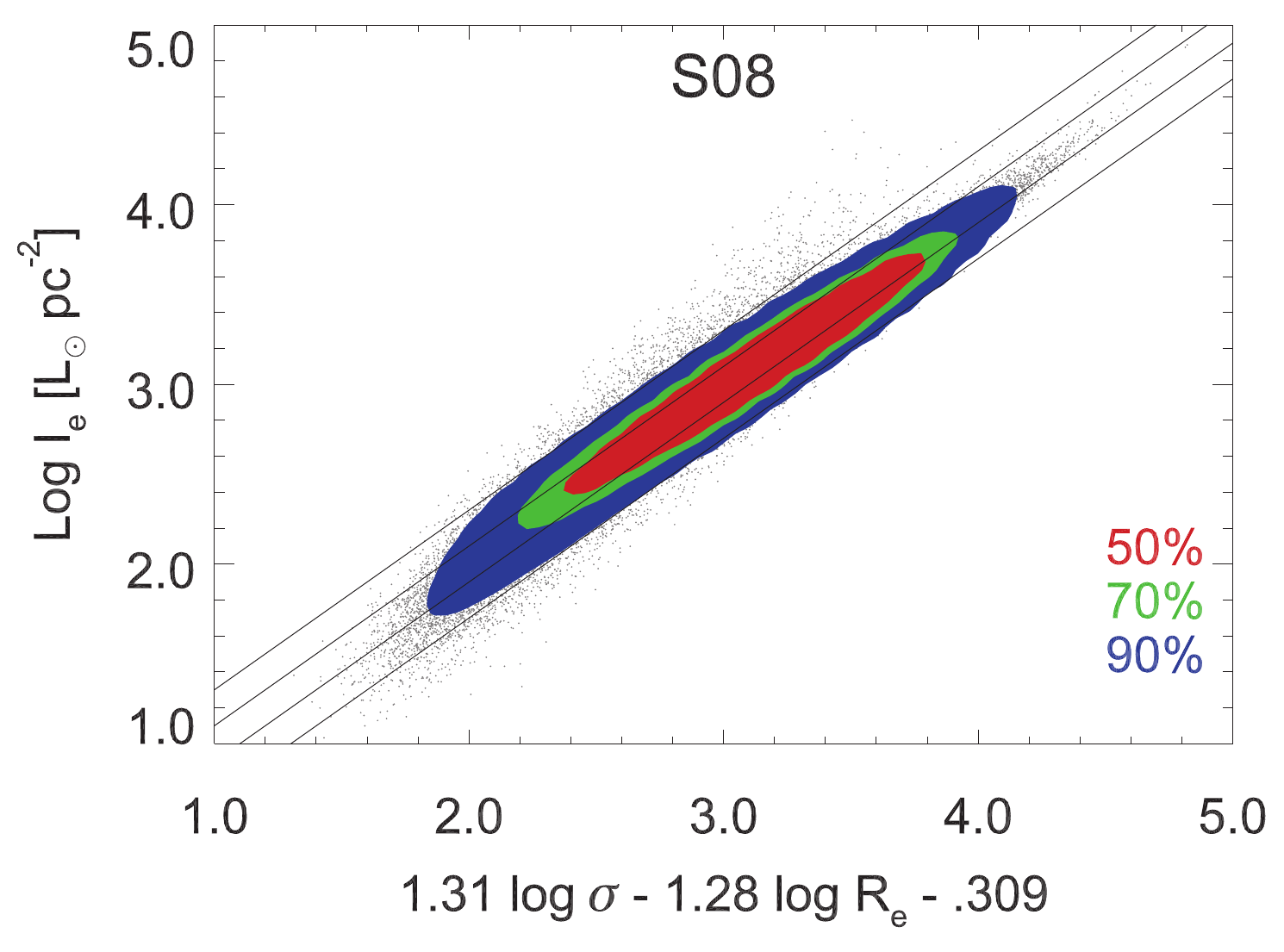}}
			{\includegraphics[width=7cm]{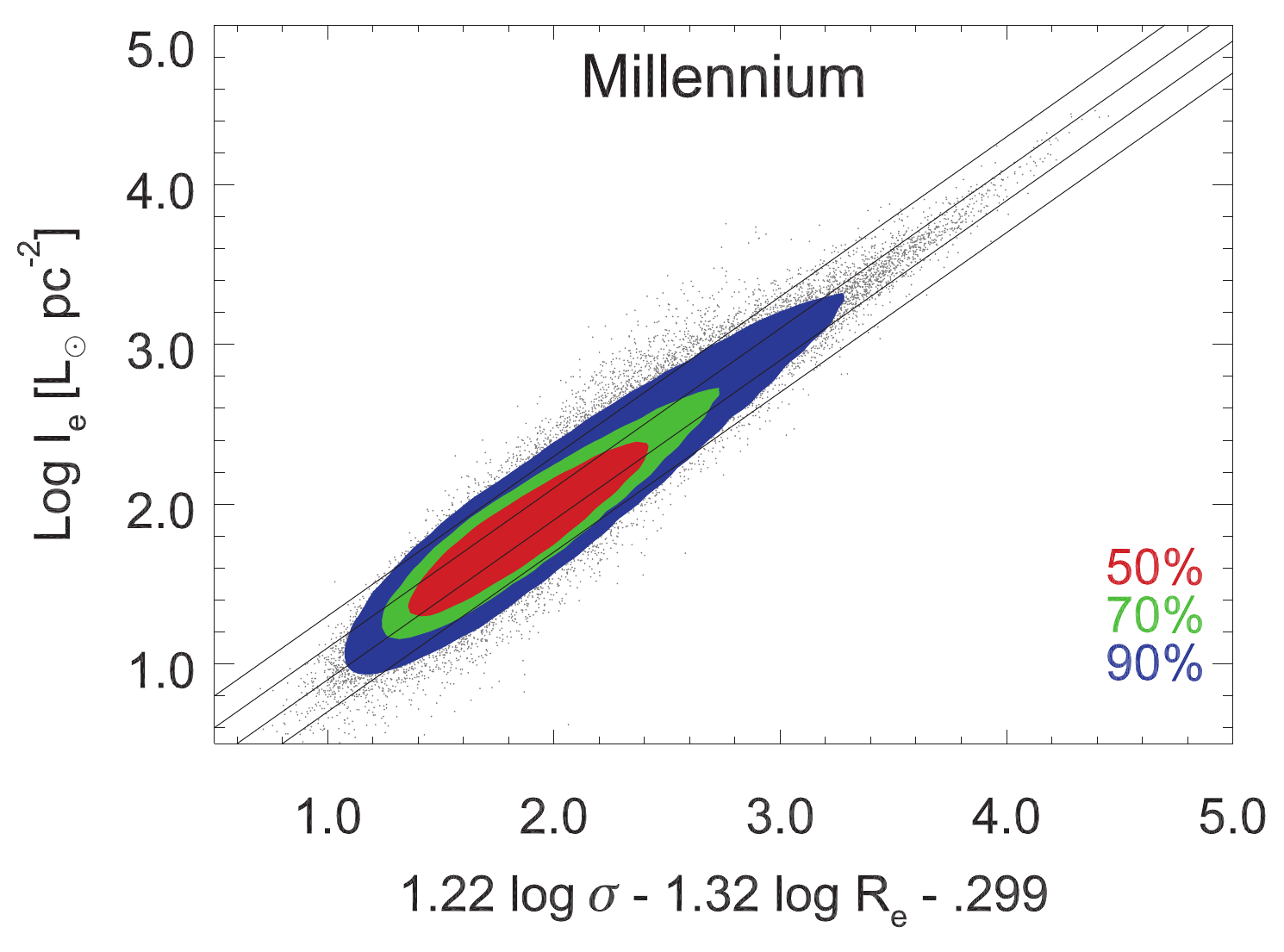}}
			}

		\caption{Distribution of galaxies through the fundamental plane for S08 (left) and Millennium (right).  Galaxies are fit to a linear relation (horizontal axis) relating surface brightness with velocity dispersion and radius.  The measured surface brightnesses are then plotted against the expected values.  The areas between the solid black lines represent the slices we term the `low-FP', `mid-FP', and `high-FP', from bottom to top, according to the residual in surface brightness. Each slice has a thickness of 0.2.  The areas below and above the red lines represent the `vlow-FP' and `vhigh-FP', respectively.  97\% and 92\% of the S08 and Millennium galaxies, respectively, fall within the middle three FP slices.  The red, green, and blue contours enclose 50\%, 70\%, and 90\% of all galaxies, while the grey points represent individual galaxies.} 
	\label{fig:FP_thickness}
\end{figure*}

If we plot the galaxies according to their location in FP-space, \SFitPercent \ (\MFitPercent) of the S08 (Millennium) galaxies fall within the low-to-high FP slices, with the majority of the outliers falling below the FP (Figure \ref{fig:FP_thickness}).

After separating the remnants by their location within the FP, we place them in bins according to their radius and velocity dispersion.  We then calculate the median age and metallicity for all galaxies within each bin, excluding bins with fewer than five galaxies for statistical purposes.  These values are used to form contours relating the stellar population parameters, namely age and metallicity, with the fundamental plane parameters and residuals.

\begin{figure*}
	\centering
		\subfigure{
			{\includegraphics[width=15cm]{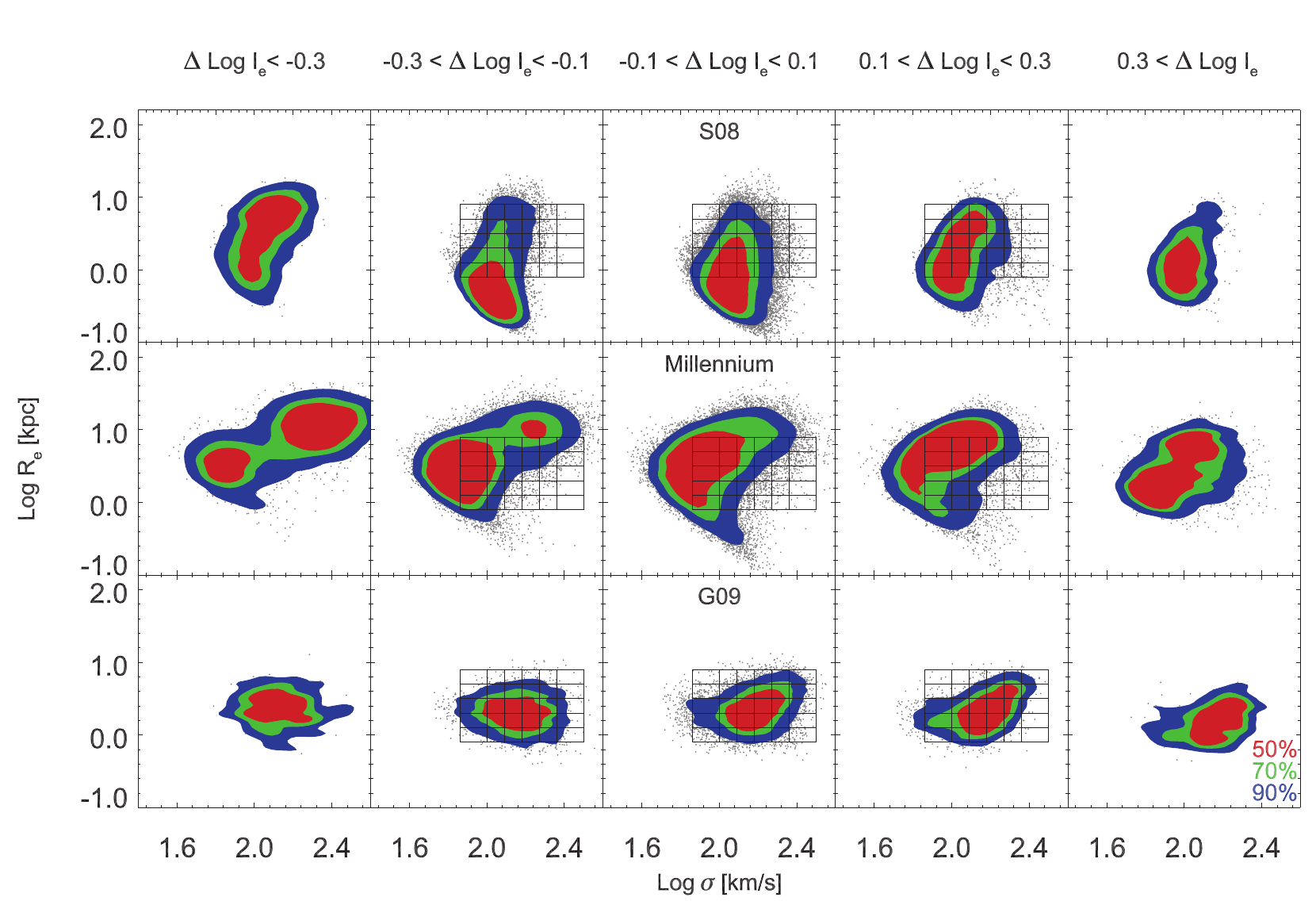}}
			}
\caption{Distribution of radius and velocity dispersion for galaxies within each slice of the FP for S08 (top), Millennium (middle), and G09 (bottom).  From left to right, the panels represent the `vlow-FP', `low-FP', `mid-FP','high-FP', and `vhigh-FP' slices.  The grid lines show the bin definitions in the region observed in G09; the median age and metallicities are calculated within each bin.  Compared to the G09 population, the Millennium galaxies are more likely to have low velocity dispersions, owing to the effects of incompleteness in the observations.  Both SAMs contain a population of galaxies with low radii and high merger redshifts; these correspond to massive, compact ellipticals and are not likely to be seen in the local universe.  The red, green, and blue contours enclose 50\%, 70\%, and 90\% of all galaxies, while the grey points represent individual galaxies.}
			\label{fig:FP_bin}
	
\end{figure*}

Comparing the S08 and Millennium populations, we find that they occupy slightly different regions of the $R_{\mathrm{e}}$-$\sigma$ parameter space (Figure \ref{fig:FP_bin}).  The Millennium galaxies tend to have larger radii and smaller velocity dispersions than the S08 galaxies.  This is a result of the differing progenitor populations, as noted in C11; Millennium disk galaxies are larger and more gas-poor than S08 disk galaxies.  As a result, major mergers induce less dissipation, producing remnants with larger radii.  Since the velocity dispersion is inversely proportional to the radius in the model, the Millennium galaxies have smaller velocity dispersions as well.  As a reference, we have overplotted the region analyzed in G09, seen as the black boxes in the middle three FP slices, as well as contours demonstrating the G09 parameter space.  Both SAMs display a population of galaxies with low radii that is not seen in G09.  These galaxies tend to have the the highest merger redshifts, suggesting that they correspond to compact ellipticals at high redshifts and are not likely to be seen in large numbers in the local universe.  The Millennium SAM also contains a population of galaxies with low velocity dispersions, $\mathrm{log \ \sigma < 1.8\ km \ s^{-1}}$, which falls below the completeness threshold of G09.  It is worth noting that the Millennium parameter space roughly resembles that of G09, with the aforementioned higher radii and lower velocity dispersions.

\section{Summary of Observations}

We compare our findings to a recent survey of early-type galaxies from the Sloan Digital Sky Survey (SDSS) \citep{York:2000a} Spectroscopic Main Galaxy Survey \citep{Strauss:2002a} Data Release 6 \citep{Adelman-McCarthy:2008a}.  The sample of galaxies is described in \cite{Graves:2009c,Graves:2009b}.  To summarize, the galaxies selected were observed in the redshift range $0.04 < z < 0.08$, with light profiles that were both centrally concentrated and fit a de Vaucouleurs profile.  To prevent a small proportion of young stars from biasing the measured luminosity, G09 excluded actively star-forming galaxies.  Using colors and emission-line intensities, G09 also rejected Seyfert hosts, low ionization nuclear emission-line region (LINER) hosts, and transition objects, as they can host active galactic nuclei (AGN) which have been found to have light profiles intermediate between early- and late-types \citep{Kauffmann:2003a}.  Although we have excluded actively star-forming galaxies from the sample, as noted above, no attempt has been made to exclude galaxies that the SAMs characterize as having active AGN at redshift zero.  

We also note that the population of simulated galaxies consists solely of ellipticals that have formed in a major merger with no subsequent evolution, while the G09 population may contain a significant number of passive S0 and Sa galaxies, as well as ellipticals that may have formed through alternative processes or undergone structural evolution following their formation.  An analysis of quiescent red sequence SDSS galaxies similar to the population studied in G09 found that 36\% of the galaxies were true elliptical galaxies, 15\% were S0 galaxies, and 48\% were Sa galaxies \citep{Cheng:2011a}.  Furthermore, while both the S08 and Millennium SAMs reproduce the early-type mass function when galaxies are separated by their stellar bulge-to-total ratios, the selection cuts we have made in this paper limit the sample to $\sim1/3$ of bulge-dominated galaxies.  Thus, a detailed comparison between the simulated galaxies and observations will require either limiting the G09 sample to pure elliptical galaxies, or expanding the simulation to include a broader range of red sequence galaxies.  This analysis will be performed in future work (Porter et al., in prep.).

Using the Lick indices \citep{Worthey:1994a,Worthey:1997a} on 16,000 stacked spectra, G09 calculated mean light-weighted ages and metallicities in bins with residual surface brightness above, within, and below the fundamental plane.  The bins covered the approximate range $\mathrm{(1.9 \ km\ s^{-1} < log \ \sigma <2.4\ km \ s^{-1})}$, $\mathrm{(0.0 \ kpc < log\ \emph{R}_{e} <0.7\ kpc)}$, and $\mathrm{(-0.3\ \Ldotpc <  \Delta\ log\ \emph{I}_{e} <0.3\ \Ldotpc)}$, where $\Delta\ \mathrm{log}\ I_{\mathrm{e}}$ is the residual surface brightness resulting from a log fit in radius and velocity dispersion, 
\begin{fleqn}
 	\begin{equation}
		\label{eqn:Graves_FP_params} 
		\mathrm{log}\ I_{\mathrm{e}} =  1.16\ \mathrm{log}\ \sigma -1.21\ \mathrm{log}\ R_{\mathrm{e}} + 0.55.
	\end{equation}
 \end{fleqn}

G09 formed contours relating the mean light-weighted age and light-weighted metallicities, [Fe/H], [Mg/H], and [Mg/Fe], to effective radius and velocity dispersion across three slices of the fundamental plane.  The Millennium and S08 SAMs measure the total metallicity \emph{Z} but do not separate contributions from different elements and only include Type II supernova; thus, we consider the total metallicity from SAMs to be most similar to [Mg/H], a measure of alpha-enhancement.  These results can be seen in Figures 7 and 9 of G09.  The authors found that stellar population age and metallicity are nearly independent of effective radius but strongly correlated with velocity dispersion.  In all three slices of the FP, galaxies with larger $\sigma$ had older ages and higher metallicities.  Stellar population age was also inversely correlated with residual surface brightness, so that the youngest galaxies tend to fall above the FP, in agreement with earlier observations \citep{Forbes:1998a,Terlevich:2002a}.

\section {Results}

\begin{figure*}
	\centering
   		\subfigure{
			{\includegraphics[width=\lbox]{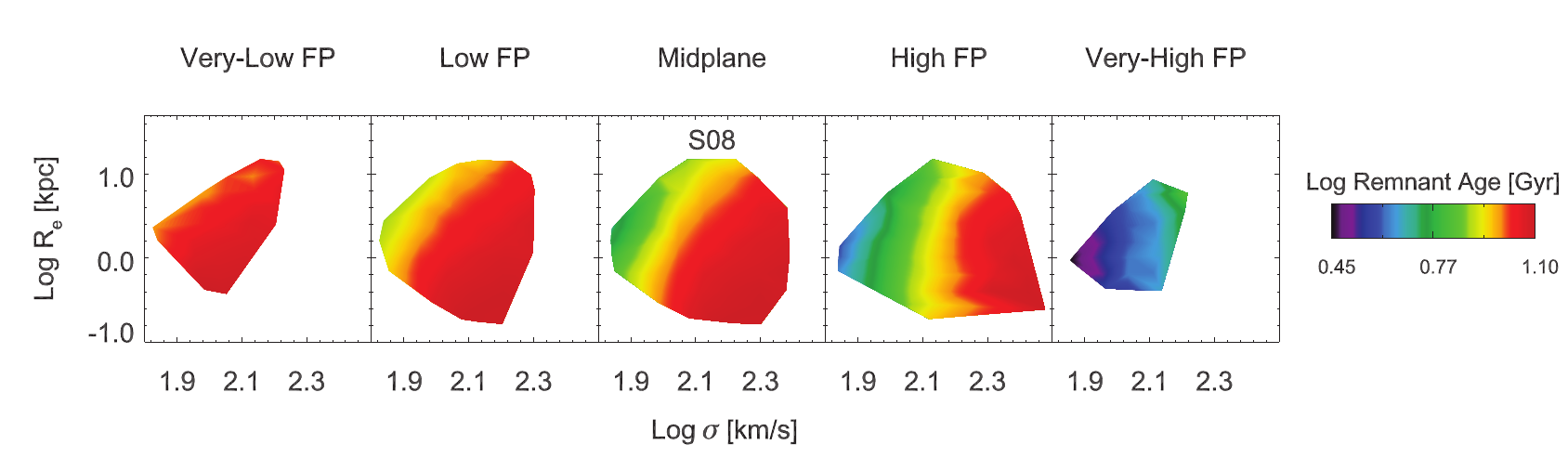}} 
			}
   		\subfigure{
			{\includegraphics[width=\lbox]{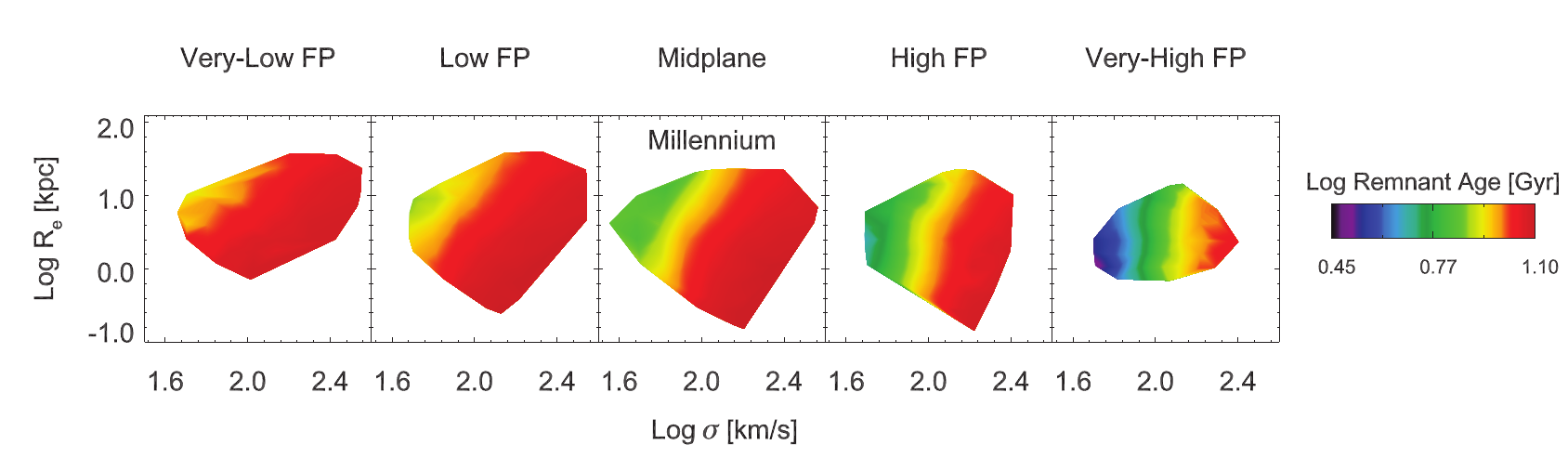}}
			}

		\caption{Relation between luminosity-weighted age, effective radius, and velocity dispersion for elliptical galaxies in S08 (top) and Millennium (bottom).  The different panels represent slices of the FP, as shown in Figure \protect\ref{fig:FP_bin}.  In both SAMs, stellar population age increases strongly with velocity dispersion and decreases with radius.  Galaxies that lie above the FP also tend to be younger than those that lie below the FP.  These results are in qualitative agreement with observations, with the exception of the dependence on radius.} 
					\label{fig:age_contour}
	
\end{figure*}

\begin{figure*}
	\centering
   		\subfigure{
			{\includegraphics[width=\lbox]{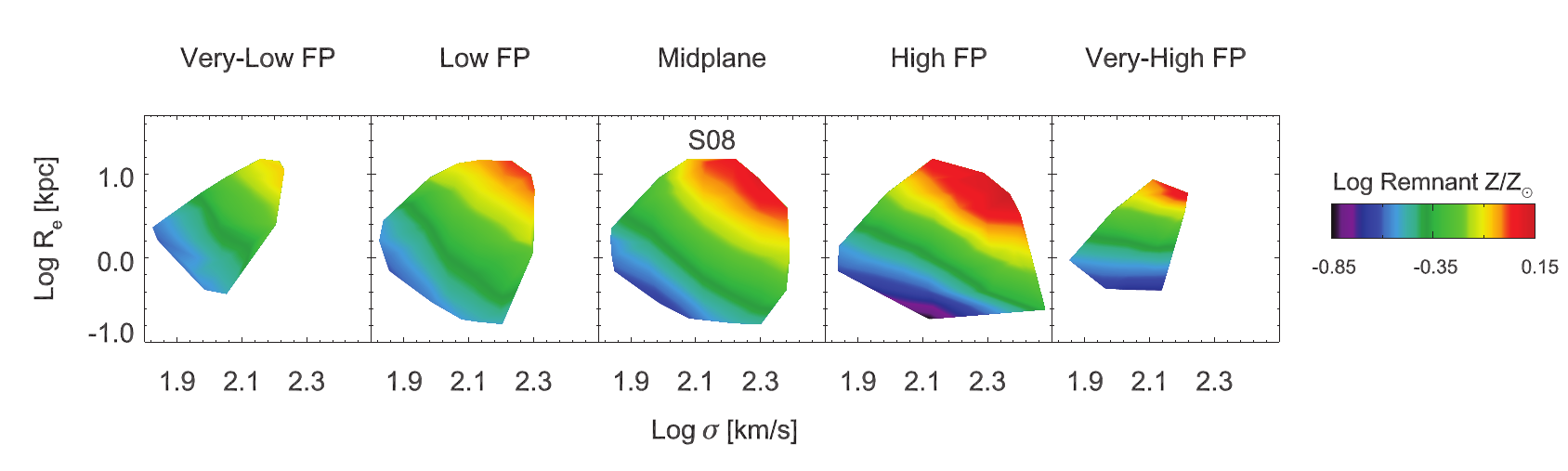}} 
  			}	
     		\subfigure{
     			{\includegraphics[width=\lbox]{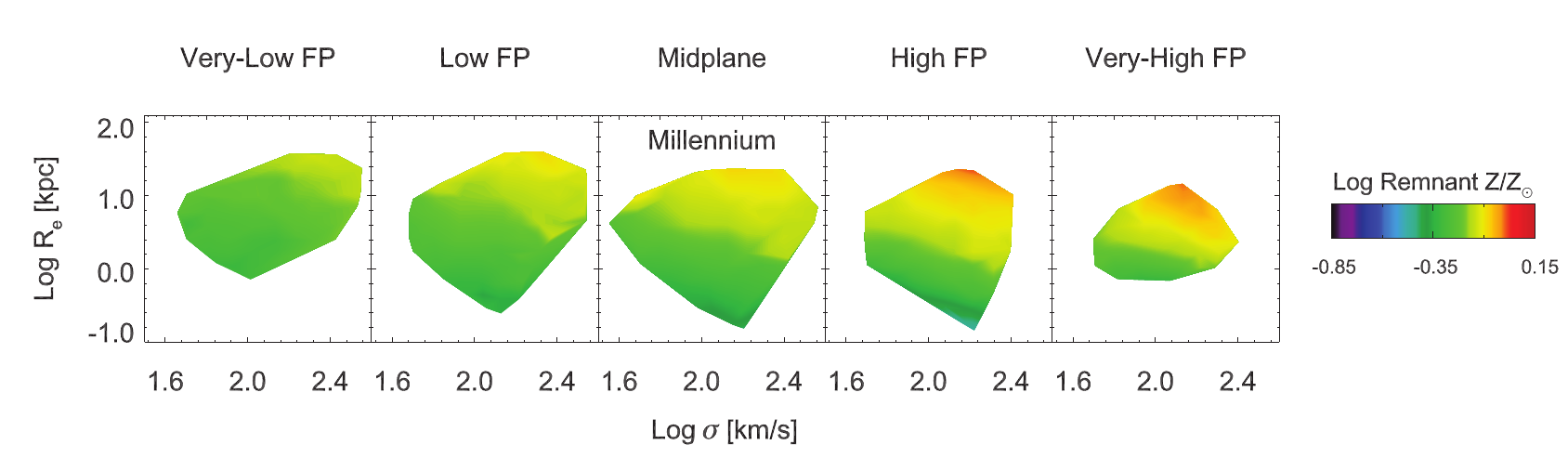}} 
			}

		\caption{Relation between luminosity-weighted metallicity, effective radius, and velocity dispersion for elliptical galaxies in S08 (top) and Millennium (bottom).  The different panels represent slices of the FP, as shown in Figure \protect\ref{fig:FP_bin}.   In both SAMs, metallicity increases with radius and velocity dispersion (contours are tilted); this is in contrast to observations, in which metallicity increases solely with velocity dispersion.  Both SAMs display a weak trend through the FP, such that galaxies that lie above the FP have slightly higher metallicities, in agreement with observations.  Galaxies from the Millennium SAM span a much lower range in metallicity.} 
				\label{fig:me_contour}
	
\end{figure*}

Our major findings are the contours seen in Figures \ref{fig:age_contour} and \ref{fig:me_contour}.  For both S08 and Millennium, stellar population age increases strongly with velocity dispersion and decreases with radius.  We note that our parameter space has a much larger range in radius than velocity dispersion, so that while the contours appear nearly vertical, the radial dependence is non-negligible.  Looking across the FP, galaxies that lie above the FP (those with the largest residuals in $\mathrm{log}\ I_{\mathrm{e}}$) have younger ages, as suggested in \cite{Graves:2009b, Forbes:1998a}; and \cite{Terlevich:2002a}.  As the Millennium galaxies sample a larger portion of the 2-dimensional $R_{\mathrm{e}}-\sigma$ parameter space (see Figure \ref{fig:FP_bin}), galaxies from the Millennium SAM tend to exhibit a larger gradient in age within a slice of the FP.

If we compare the metallicity contours, the differences between the SAMs are slightly more pronounced.  In both cases in Figure 4, metallicity increases with velocity dispersion and radius.  However, the Millennium remnants have a stronger correlation with radius, while for S08 the trends with radius and velocity dispersion are roughly equal.  Galaxies that lie above the FP tend to have higher metallicities, as in G09.

\begin{figure*}
	\centering
   		\subfigure{
			{\includegraphics[width=8cm]{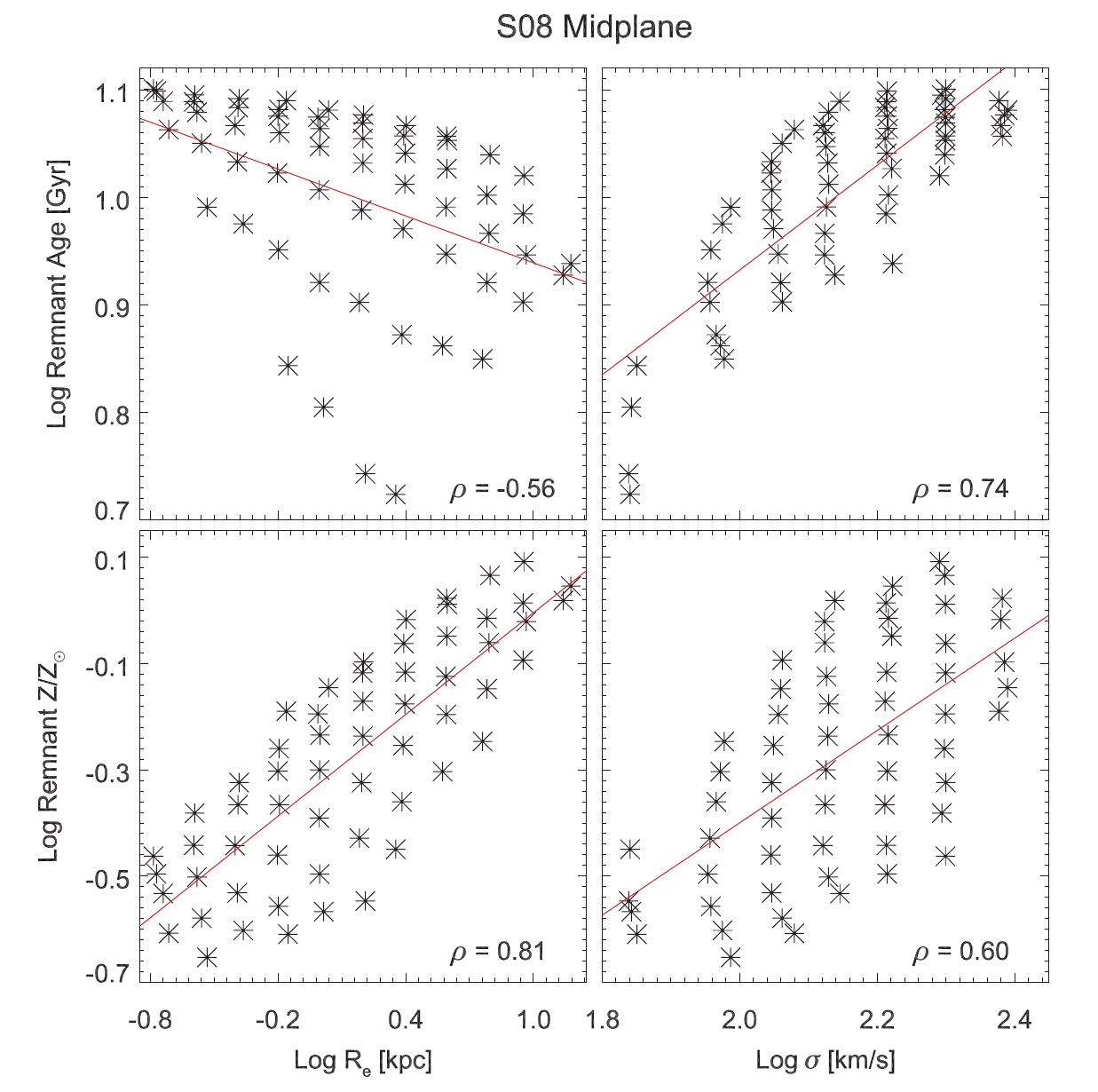}} 
  			}	
     		\subfigure{
     			{\includegraphics[width=8cm]{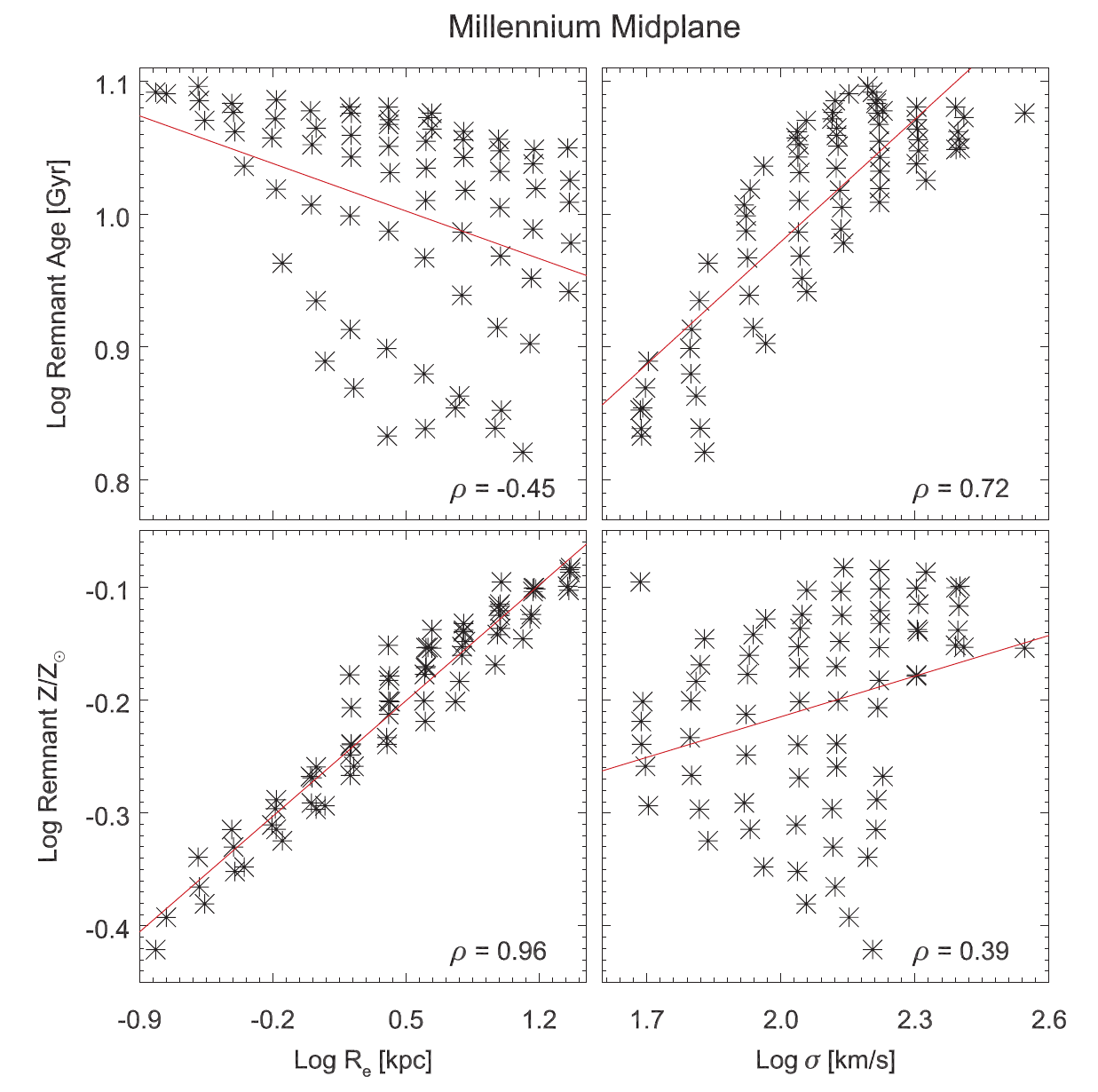}} 
			}

		\caption{Dependence of stellar age and metallicity on radius and velocity dispersion for elliptical galaxies in S08 (left) and Millennium (right).  Each panel represents the midplane of the FP, as shown in Figure \protect\ref{fig:FP_bin}.  The solid red line represents a least-squares fit to the binned data. $\rho$ is the Spearman rank coefficient for the binned data, where $\rho = 1.0 $ designates a perfect correlation, $\rho = -1.0 $ designates a perfect anti-correlation, and $\rho = 0.0$ designates no correlation.  Both SAMs demonstrate a strong correlation between age and velocity dispersion ($\rho = 0.74$ and 0.72 for S08 and Millennium, respectively), and a somewhat weaker anti-correlation between age and effective radius ($\rho = -0.56$ and -0.45 for S08 and Millennium, respectively).  The Millennium metallicity has a very tight dependence on effective radius ($\rho = 0.96$) and a much weaker dependence on velocity dispersion ($\rho=0.39$), while the S08 metallicity has a moderate dependence on both radius and velocity dispersion ($\rho = 0.81$ and 0.60, respectively).}
				\label{fig:FP_projected}
	
\end{figure*}

To better quantify the dependence of the stellar population parameters on the FP parameters, we have plotted the binned stellar age and metallicity as a function of radius and velocity dispersion for the midplane of both SAMs (Figure \ref{fig:FP_projected}).  We have also calculated the least-squares fit and Spearman rank coefficients for these relations.   These two-dimensional relations confirm the trends seen in the contours.  Both SAMs have a stronger dependence on velocity dispersion ($\rho = $ 0.74 and 0.72 for S08 and Millennium, respectively) than on effective radius ($\rho = $ -0.56 and -0.45 for S08 and Millennium, respectively) with regards to stellar age.  The Millennium metallicities have a nearly perfect correlation with radius ($\rho = 0.96$) and a weak correlation with velocity dispersion ($\rho = 0.39$), while the S08 metallicities have a somewhat stronger correlation with radius ($\rho = 0.81$) than with velocity dispersion ($\rho = 0.60$).  With the exception of the Millennium metallicity-radius relation, all of the correlations have a large degree of scatter around the least-squares fit.

There are marked differences between the ranges of metallicities in S08 and Millennium.  Comparing the 25th and 75th percentiles, the range of metallicites in S08 ($ -0.50 < \log Z < -0.24 $) is roughly twice that of Millennium ($-0.26 < \log Z < -0.14$).  While the average metallicity in S08 is lower than in Millennium ($\log Z = -0.37 $ and $ -0.20$, respectively), the overall gradient in metallicity is also much larger in S08, with the low (high) $\sigma$ galaxies having lower (higher) metallicities than in Millennium.

We defer a detailed comparison between the two SAMs to future work, but we note here that the two SAMs use similar prescriptions for metal enrichment, modeled after \cite{De-Lucia:2004b}; for every solar mass of star formation, a fixed yield of metals is returned to the cold gas disk.  There are however key differences between the two SAMs.  The S08 progenitor galaxies span a larger range in metallicity as a function of stellar mass (see Figure \ref{fig:Prog_mass_age_me}, right pane).  Furthermore, the average stellar mass of a remnant is lower in S08 than Millennium, with values of $\log M_{\odot} = 10.37 $ and $ 10.52$, respectively.  Finally, the S08 galaxies tend to merge at higher redshifts than their Millennium counterparts, with the average merger redshift being 1.73 and 0.99, respectively.  Since metallicity tends to increase with stellar mass \citep{Gallazzi:2005a} and decrease with redshift \citep{Pipino:2011a,Laskar:2011a,Calura:2009a,Maiolino:2008a,Pipino:2008a,Erb:2006b}, the latter two trends bias the S08 galaxies toward lower metallicities.  However we note that, even at fixed redshift, the mass-metallicity has a larger gradient in S08 than in Millennium (see Figure \ref{fig:Prog_mass_corrs_z}, right panel).

If we neglect the range in metallicity within the SAMs, we find that the two SAMs produce remarkably similar age-FP and metallicity-FP relations.  As we will explain in more detail later, this agreement stems from the correlations between the progenitor gas fraction, age, and metallicity, on the one hand, and the progenitor stellar mass, on the other.  Since the progenitor populations of the two SAMs have similar stellar mass dependences, the remnants have similar fundamental plane relations.

Stellar population trends through the thickness the FP can arise in a number of different ways; any process that increases the dynamical mass-to-light ratio of a galaxy will move it further below the virial plane.  \cite{Graves:2010a} provide a decomposition of the deviation from the virial theorem, separating it into four components: 
\begin{enumerate}
\item The ratio of the estimated dynamical mass to the true mass within one $\mathrm{R_e}$.
\item The ratio of the true mass within one $R_e$ to the projected stellar mass.
\item The ratio of the projected stellar mass with an assumed initial mass function (IMF) to the true projected stellar mass and its corresponding IMF.
\item The stellar mass-to-light ratio for the assumed IMF.
\end{enumerate}  For the simulated galaxies and their corresponding FP, the first and third terms are identically one, as we have no uncertainty in the dynamical mass estimate and we model and `observe' galaxies using the same IMF.  Since we know the stellar mass-to-light ratio for the galaxies and can calculate the central dark matter fraction (DMF, see below), we may calculate the second and fourth terms directly.

\begin{figure*}
	\centering
   		\subfigure{
			{\includegraphics[width=\lbox]{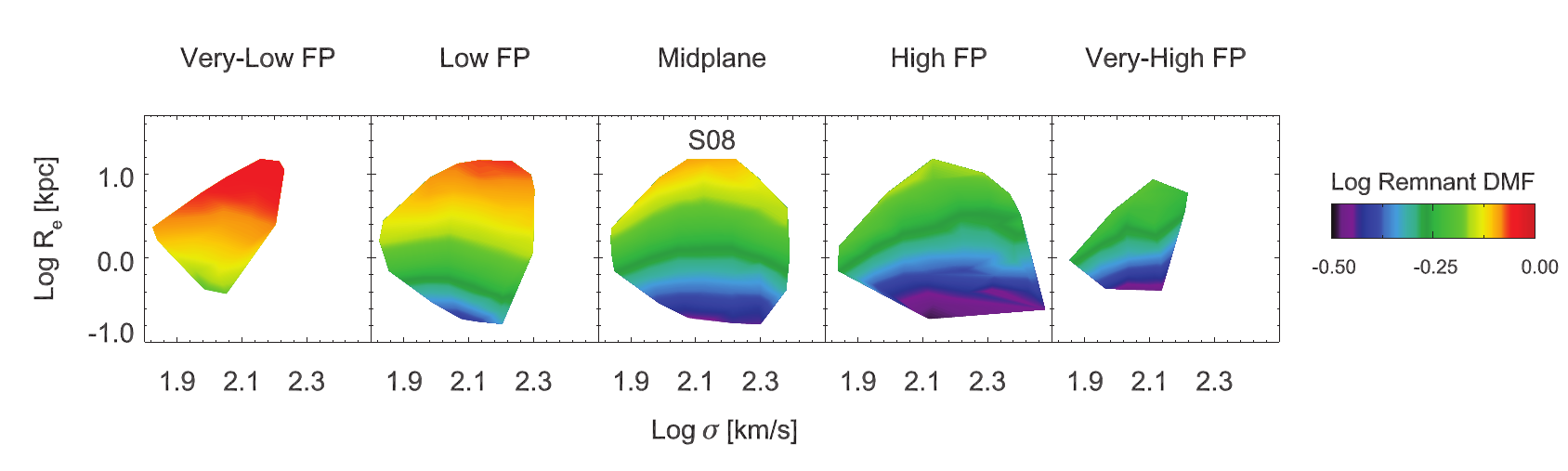}} 
			}
   		\subfigure{
			{\includegraphics[width=\lbox]{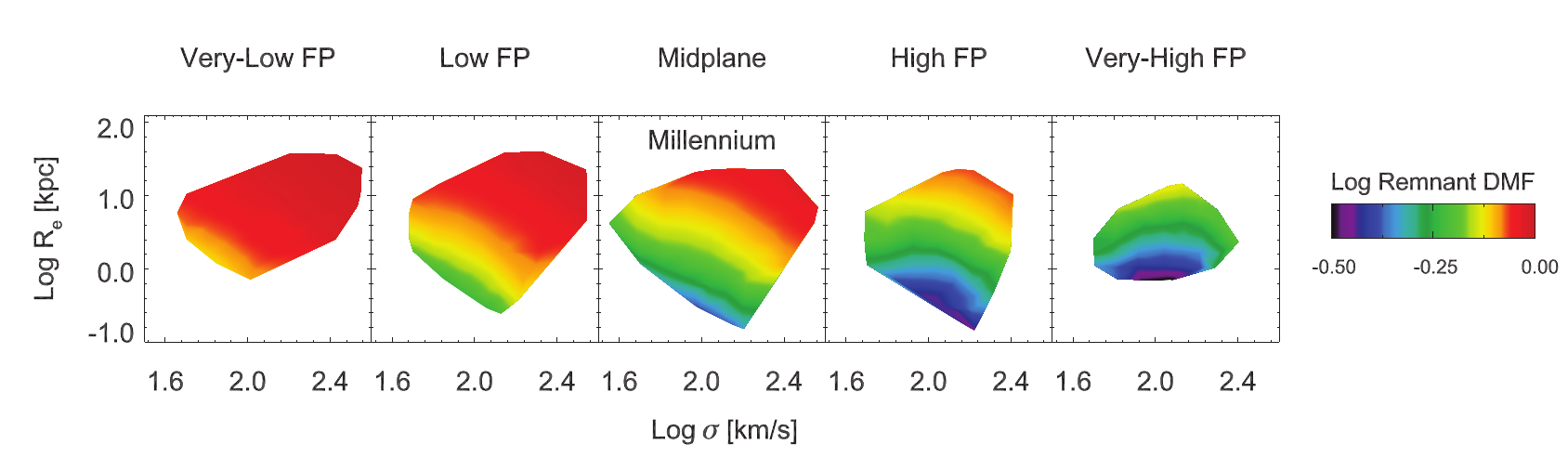}}
			}

		\caption{Relation between central dark matter fraction, effective radius, and velocity dispersion for elliptical galaxies in S08 (top) and Millennium (bottom).  The different panels represent slices of the FP, as shown in Figure \protect\ref{fig:FP_bin}.  In both SAMs, galaxies that fall below the FP have higher central dark matter fractions.} 
					\label{fig:dmf_contour}
	
\end{figure*}

\begin{figure*}
	\centering
   		\subfigure{
			{\includegraphics[width=\lbox]{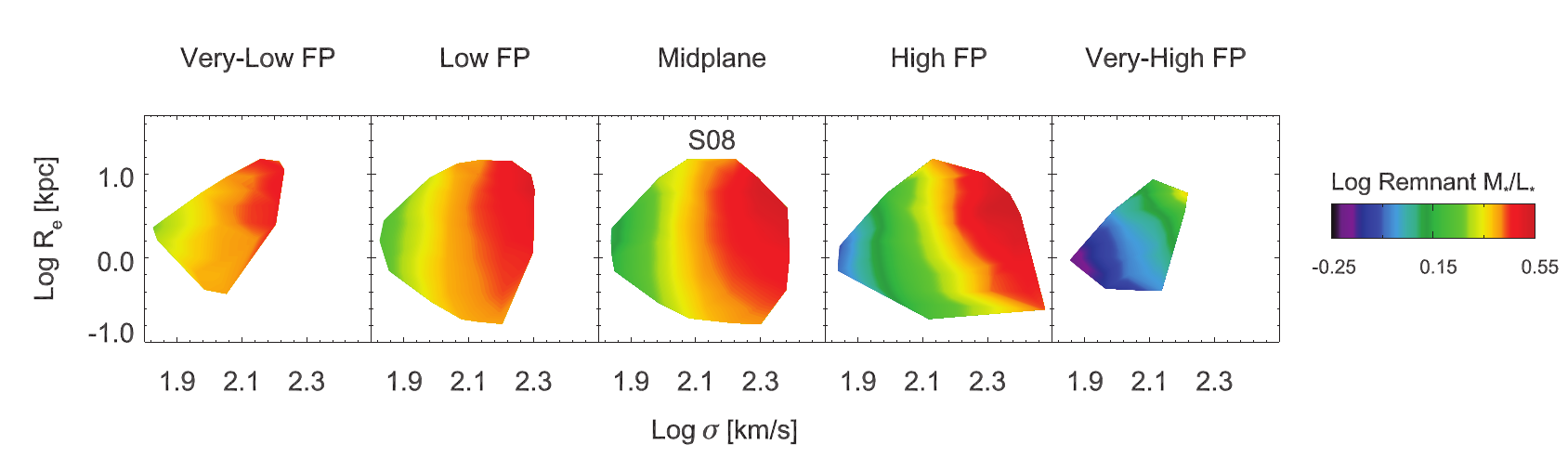}} 
  			}	
     		\subfigure{
     			{\includegraphics[width=\lbox]{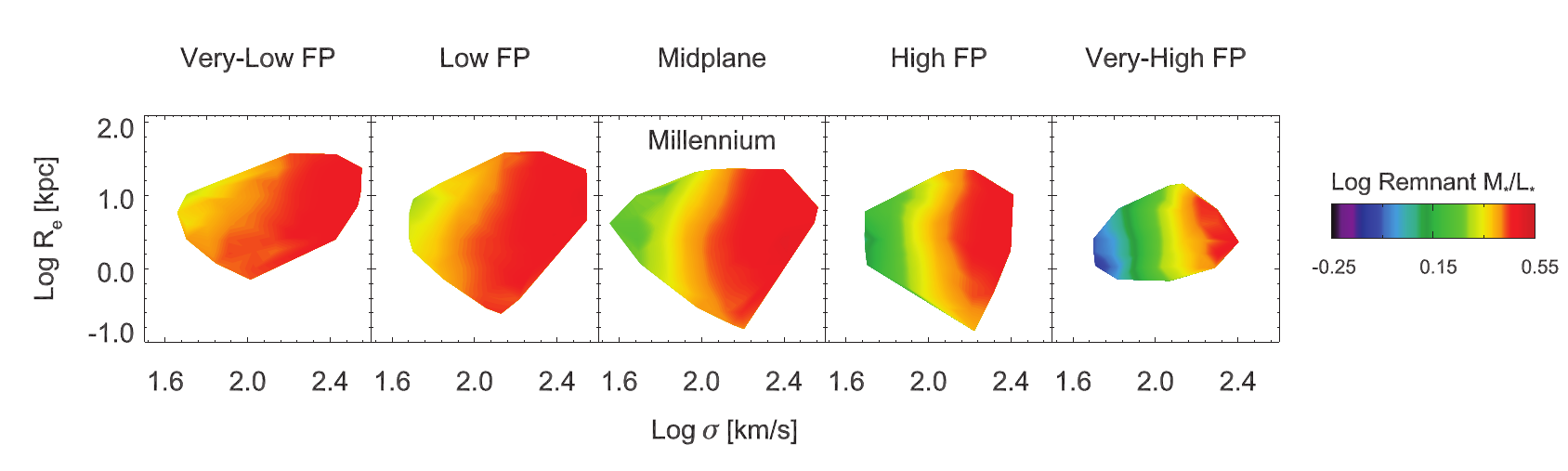}} 
}

		\caption{Relation between stellar mass-to-light ratio, effective radius, and velocity dispersion for elliptical galaxies in S08 (top) and Millennium (bottom).  The different panels represent slices of the FP, as shown in Figure \protect\ref{fig:FP_bin}.  Galaxies that lie above the FP tend to have lower mass-to-light ratios} 
				\label{fig:mtol_contour}
	
\end{figure*}

We have used the same process as described above to project the central dark matter fractions and stellar mass-to-light ratios of elliptical galaxies across and through the FP (Figures \ref{fig:dmf_contour} and \ref{fig:mtol_contour}).  The results are in agreement with the conclusions of \cite{Graves:2010a}: galaxies that fall below (above) the FP have higher (lower) dark matter fractions and higher (lower) stellar mass-to-light ratios.  Furthermore, the dark matter fraction tends to increase with radius, while the mass-to-light ratio increases with velocity dispersion.  The former result is not surprising, as the ratio of dark matter to baryonic matter increases with radius; thus, larger galaxies enclose more dark matter.  Comparing the trends through the thickness of the FP, at fixed $\mathrm R_{e}$ and $\sigma$ the dark matter fraction has a larger degree of variation than the mass-to-light ratio.  If we limit the sample to the middle three FP planes analyzed in G09 and \cite{Graves:2010a}, we find that the average variance in the dark matter fraction contributes 80\% and 88\% of the thickness of the FP for S08 and Millennium, respectively.  This is in general agreement with \cite{Graves:2010a}, who found that the dark matter fraction and IMF variation have a combined contribution in the range of 47\% - 98\%.  This is an indication that underlying structural differences, as opposed to stellar population effects, are the main contributors to the thickness of the FP.

The variation in the central dark matter fraction can arise in a variety of ways: an increase in the stellar mass within one effective radius, an increase in the concentration of the host halo, and an increase in the effective radius at a given stellar mass will all have the effect of increasing the central dark matter fraction.
In the dissipational model, the central dark matter fraction is estimated by calculating the halo half-mass radius, assuming that the two progenitor halos merge dissipationlessly, and fitting the resulting radius to an isothermal profile to calculate the mass of dark matter within one stellar effective radius.  At fixed mass and stellar half-mass radius, halos with larger dark matter half-mass radii will have lower concentrations and lower central dark matter fractions.  We choose an isothermal profile, as both simulations and observations have suggested that the central regions of dark matter halos containing early-type galaxies are isothermal in nature \citep{Auger:2010a,Auger:2010b,Bolton:2008a,Duffy:2010a,Koopmans:2006a}. For a dissipationless merger, 
\begin{fleqn}
	\begin{equation}
R_{\mathrm{dm,f}}=\frac{(M_\mathrm{{dm,1}}+M_\mathrm{{dm,2}})^2}{\frac{M_\mathrm{{dm,1}}^2}{R_\mathrm{{dm,1}}}+\frac{M_\mathrm{{dm,2}}^2}{R_\mathrm{{dm,2}}}}
	\end{equation} 
 \end{fleqn}
where ${M_\mathrm{{dm,1}}}$ and ${M_\mathrm{{dm,2}}}$ are the total masses of the dark matter haloes of the two progenitors,  ${R_\mathrm{{dm,1}}}$ and ${R_\mathrm{{dm,2}}}$ are the half-mass radii of the haloes of the progenitors, and  ${R_\mathrm{{dm,f}}}$ is the halo half-mass radius of the remnant.  Progenitors with small halo half-mass radii will produce a remnant with a correspondingly contracted halo and a higher concentration of dark matter within one $R_\mathrm{e}$.  Thus, the variation of the dark matter fraction at fixed $R_{e}$ through the FP arises from variations in the concentrations of the progenitor spiral galaxies. 

\begin{figure*}
	\centering
   		\subfigure{
			{\includegraphics[width=10cm]{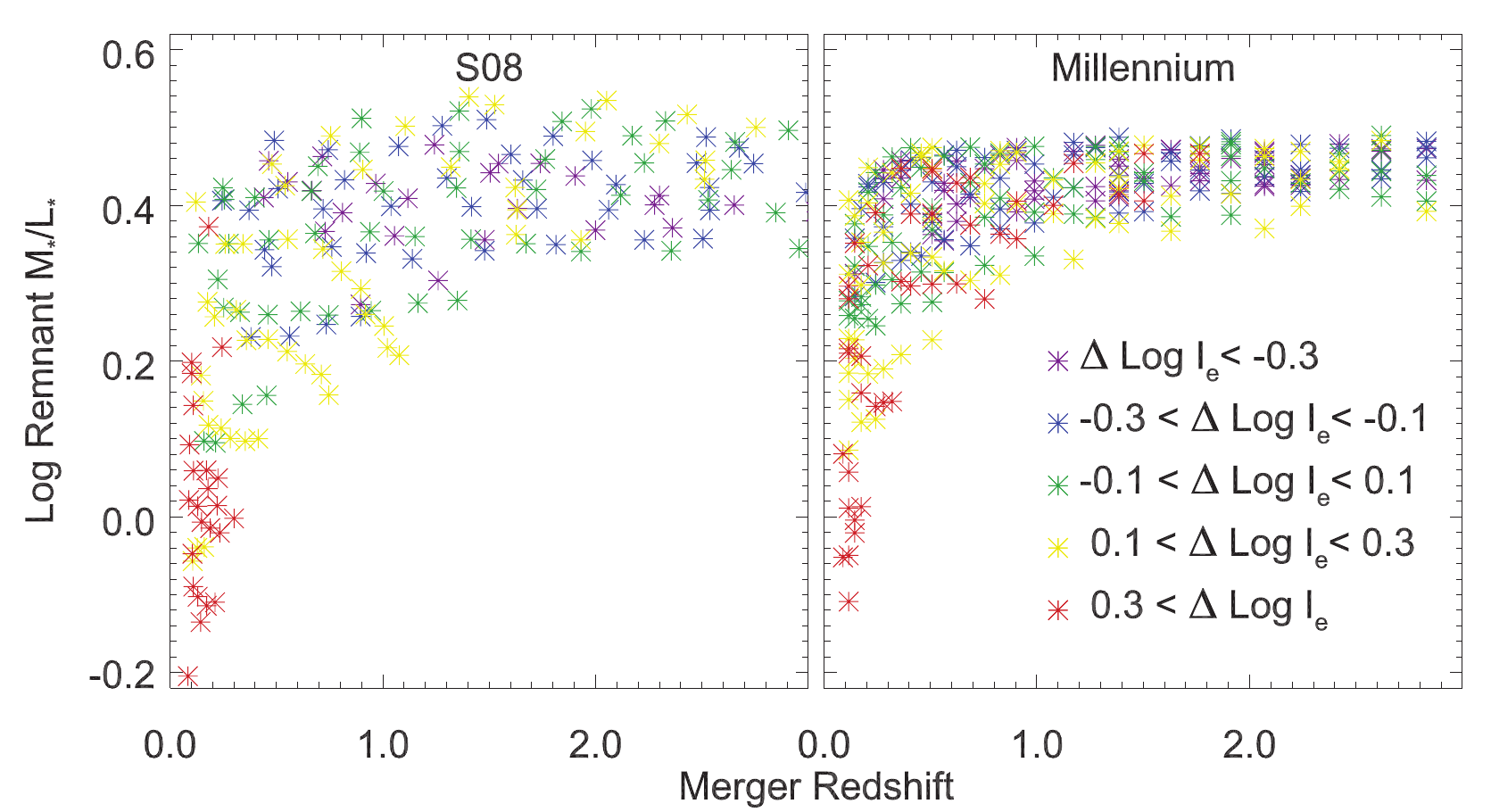}} 
  			}	
		\caption{Relation between stellar mass-to-light ratio and the redshift of the merger for galaxies in S08 (left) and Millennium (right).  Different colors represent different slices of the FP, as shown in Figure \protect\ref{fig:FP_bin}.  While galaxies that have very low merger redshifts tend to lie in the highest FP slice and have low mass-to-light ratios, there is no correlation between merger redshift and mass-to-light ratio in the other FP slices.} 
				\label{fig:mtol_redshift}
	
\end{figure*}

The variation in the stellar mass-to-light ratio can further be separated into two components.  Galaxies that merge at lower redshifts will have more recent star formation and decreased mass-to-light ratios, but there may also be an intrinsic variation in galaxies at a fixed redshift due to differences in stellar age.  When we calculate the median mass-to-light ratio and merger redshift within each bin in $R_{\mathrm{e}}$ and $\sigma$, and compare these values through the FP, we find no correlation between merger redshift and stellar mass-to-light ratio in the lower four FP slices, while the vast majority of the galaxies in the highest FP slice have low merger redshifts and low mass-to-light-ratios (Figure \ref{fig:mtol_redshift}).  This suggests that the effects of recent mergers are limited to galaxies that lie far above the FP, and that variations within the other four slices represent intrinsic differences.  This is consistent with \cite{Graves:2010a}, who found that the thickness of the FP was due to structural differences in the stellar mass surface density rather than the passive fading of galaxies with young stars.

\subsection{Role of progenitor scaling relations}
\label{ssec:Progenitors}
Since our work is based on merging late-type galaxies with properties taken from the Millennium and S08 SAMs, the results will reflect any biases present in the SAM populations.  Comparing the evolution of the Millennium progenitors' size-mass relations to observational data, C11 found that spiral galaxies that merged at the lowest redshifts, 0.0 $< $ z $<$ 0.5, were on average 50\% too large, while the most massive galaxies were too large at all redshifts.  The slope of the late-type size-mass relation, however, was consistent with observational results, both in the local universe and at high redshifts.  The size-mass relation for elliptical galaxies following a merger obeyed the same trends, with a smaller radius offset. As a result, the Millennium galaxies shown here have larger radii than observations would suggest.  The S08 progenitors and remnants are, in general, more consistent with observations, though remnants with the lowest formation redshifts are systematically large by 0.2 dex.

To determine the effect of the major merger itself on the age-FP and metallicity-FP relations, we have formed similar contours using the properties of the disk progenitors.  Using the same methodology as described above, progenitors are placed along a plane relating surface brightness, effective radius, and circular velocity.  \cite{Burstein:1997b} found that the radii, surface brightnesses, and characteristic velocities of self-gravitating systems form a `cosmic metaplane', where the slope of the respective plane varies with Hubble type.  For our purposes, we we will call the late-type portion of this metaplane the late-type fundamental plane (LTFP).   

For consistency, we use the surface brightness that the progenitor galaxies would have had at redshift zero if they had not undergone the merger event and all star formation stopped at the time of the merger.  While we do not expect this to be entirely accurate, we make this approximation in order to compare to the low-redshift results of G09.  We note that LTFP and FP relations across a single slice are not directly related, since progenitors that lie below the FP may produce remnants that lie above it.  The overall trends, however, are consistent across each LTFP slice, allowing us to make comparisons between trends in $R_{\mathrm{e}}$ and $V_{circ}$ for the progenitors and those in $R_{\mathrm{e}}$ and $\sigma$ for the remnants.

\begin{figure*}
	\centering
   		\subfigure{
			{\includegraphics[width=\lbox]{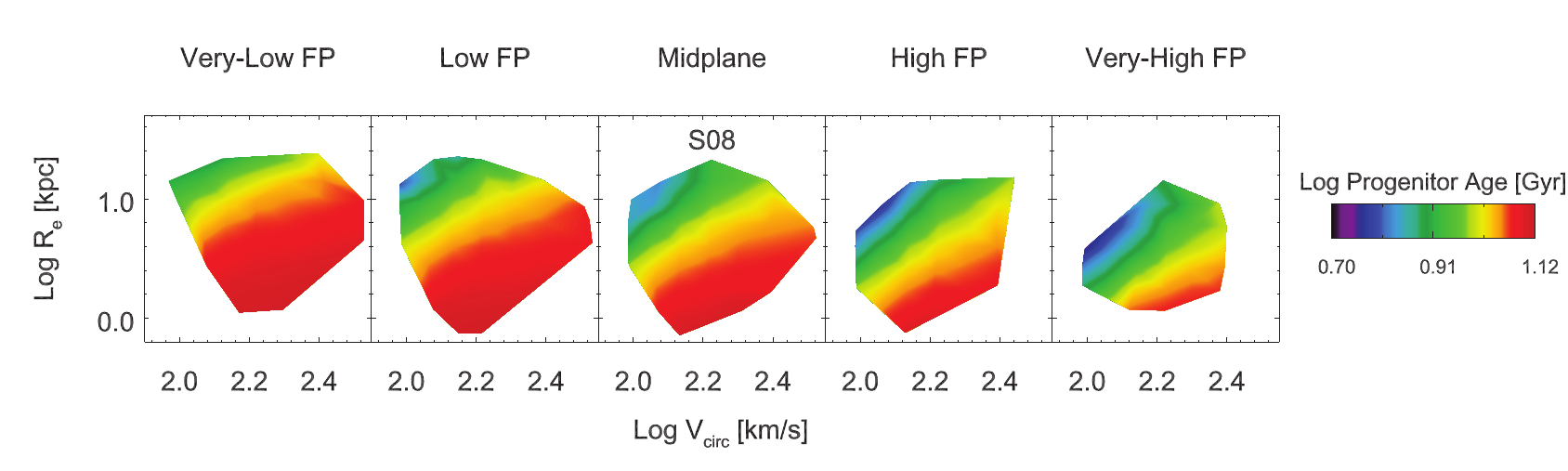}} 
			}
   		\subfigure{
			{\includegraphics[width=\lbox]{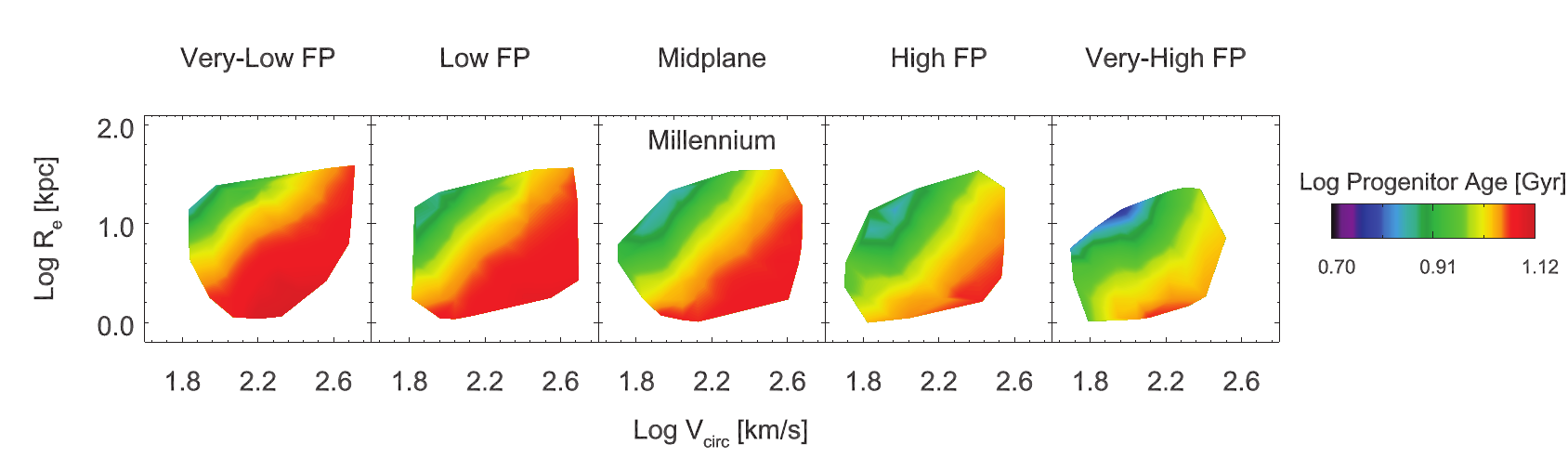}}
			}

		\caption{Relation between luminosity-weighted age, effective radius, and circular velocity for spiral progenitor galaxies in S08 (top) and Millennium (bottom).  The different panels represent slices of the late-type fundamental plane (LTFP). In both SAMs, stellar population age increases with circular velocity and decreases with radius.} 
						\label{fig:age_prog}
	
\end{figure*}

\begin{figure*}
	\centering
   		\subfigure{
			{\includegraphics[width=\lbox]{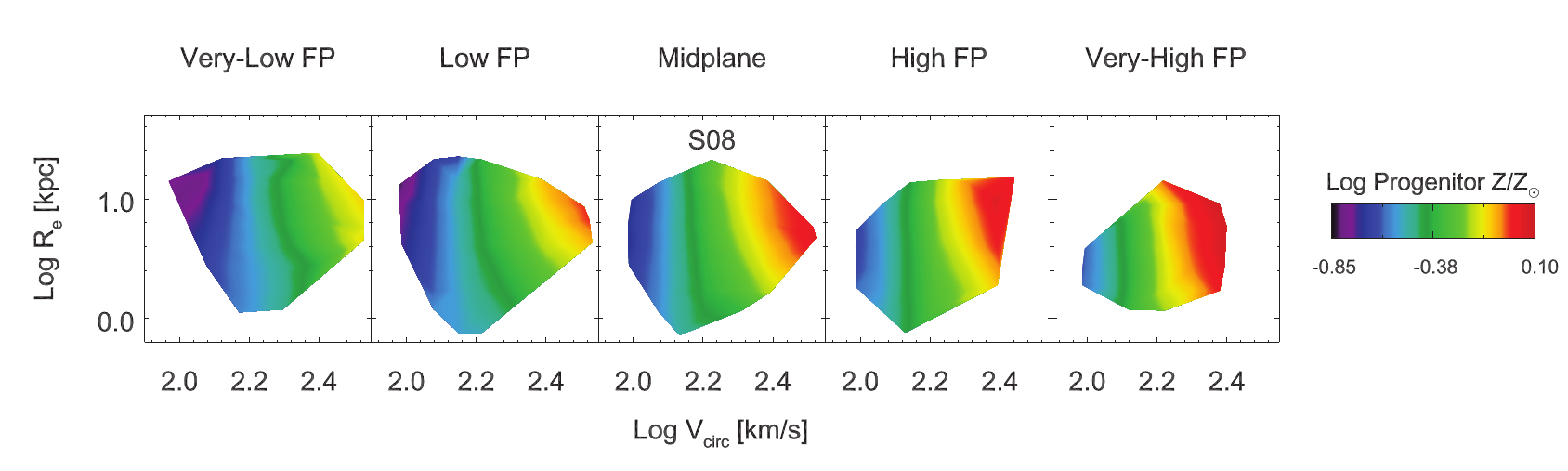}} 
  			}	
     		\subfigure{
     			{\includegraphics[width=\lbox]{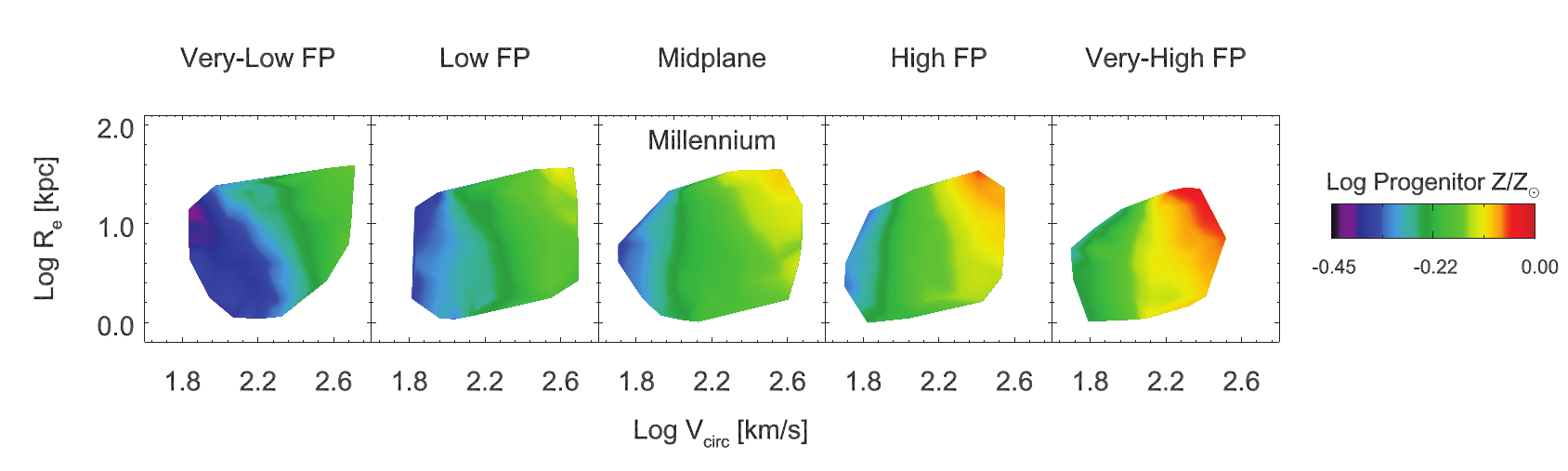}} 
			}

		\caption{Relation between luminosity-weighted metallicity, effective radius, and circular velocity for spiral progenitor galaxies in S08 (top) and Millennium (bottom).  The different panels represent slices of the LTFP. In both SAMs, metallicity increases with circular velocity but is nearly independent of radius.} 
		 			\label{fig:me_prog}
	
\end{figure*}

Results are shown in Figures \ref{fig:age_prog} and \ref{fig:me_prog}.  Overall, we find that progenitor ages increase with circular velocity and decrease with radius, while their metallicities increase strongly with circular velocity and are nearly independent of radius.

\begin{figure*}
	\centering
   			{\includegraphics[width=\lbox]{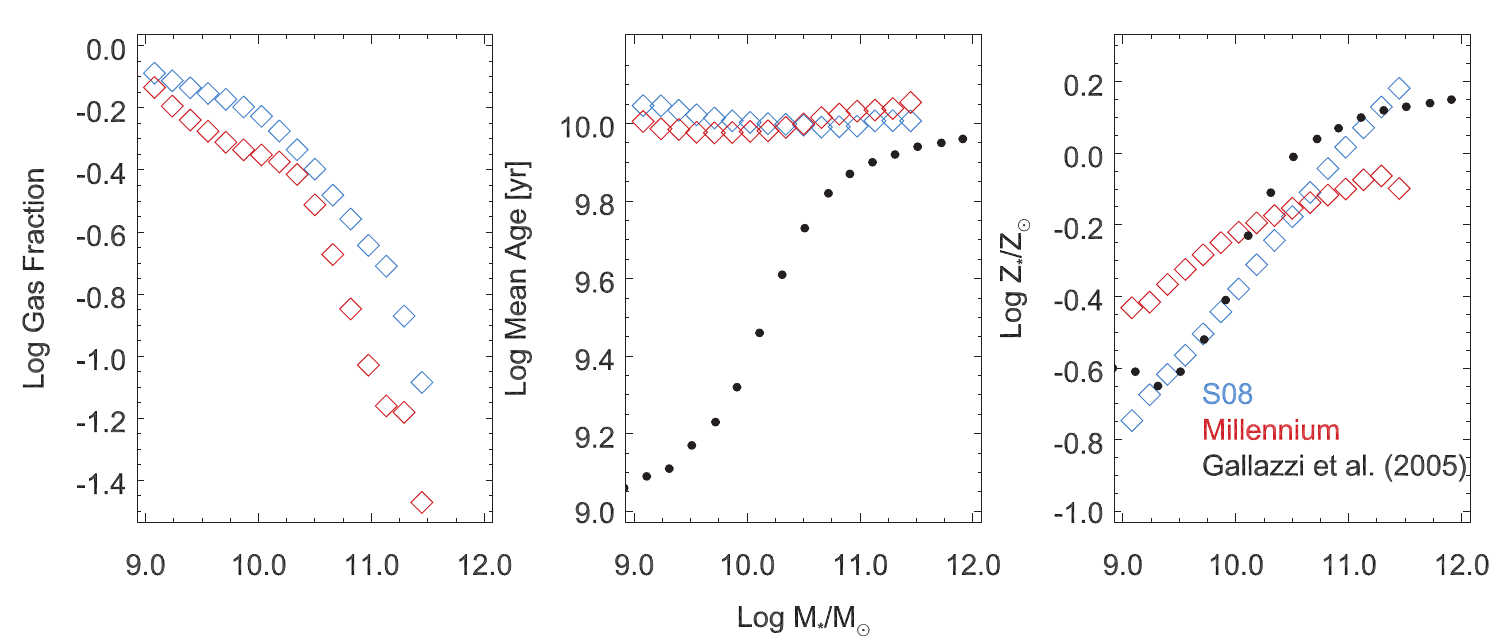}} 
     				\caption{From left: stellar gas fraction, age, and metallicity as a function of stellar mass for spiral progenitors from the S08 (blue diamonds) and Millennium (red diamonds) SAMs. The black circles represent SDSS observations from \protect\cite{Gallazzi:2005a}.  In both SAMs, gas fraction and metallicity decrease with stellar mass in qualitative agreement with observations, though the Millennium progenitors span a much smaller range in metallicity than their S08 counterparts.  However, both SAMs produce a flat age-mass relation, in disagreement with observations.}
   			\label{fig:Prog_mass_age_me}
	
\end{figure*}

The differing age-FP and metallicity-FP relations for the progenitors stem from differences in the age-mass and metallicity-mass relations in the two models.  Observations have shown that more massive spiral and elliptical galaxies tend to have older ages and higher metallicities than their less massive counterparts \citep{Gallazzi:2005a}.  While both the Millennium and S08 progenitors have increasing metallicities as a function of mass, their ages are nearly independent of mass (Figure \ref{fig:Prog_mass_age_me}, see also \cite{Somerville:2008b}, Figure 6).  Since the stellar mass is related to the circular velocity via the Tully-Fisher relation \citep{Tully:1977a}, $M_{*} \propto V_{\rm circ}^{\alpha}$, $\alpha \sim 4$, the age-mass and metallicity-mass correlations directly translate into age-LTFP and metallicity-LTFP relations for the progenitors; both age and metallicity should increase with circular velocity while remaining relatively independent of radius.  

The failure of the SAMS to reproduce the age-mass relation can be seen clearly in Figure \ref{fig:age_prog}, as the radial dependence of the stellar age is in the opposite direction from what the age-mass relation would predict.  Instead of a strong age-mass relation, we find that the progenitors' ages are linked to the redshift of the merger.  Since we calculate all stellar properties at redshift zero but truncate star formation at the time of the merger, galaxies that merge at high redshifts have intrinsically older ages than those that merge at lower redshifts.  This represents an artifact of the analysis methods, which will be corrected when we include subsequent evolution in future work.  As we will show in section 5, however, the age-FP and metallicity-FP relations are similar regardless of the redshift of the merger.

As the SAMs correctly reproduce the metallicity-mass relation for spiral galaxies, they predict a metallicity-LTFP relation that is strongly dependent on circular velocity and is nearly independent of radius (Figure \ref{fig:me_prog}).  It is also worth noting that both the Millennium progenitors and remnants span a much smaller range in metallicity than their S08 counterparts.  

\begin{figure*}
	\centering
   		\subfigure{
			{\includegraphics[width=8cm]{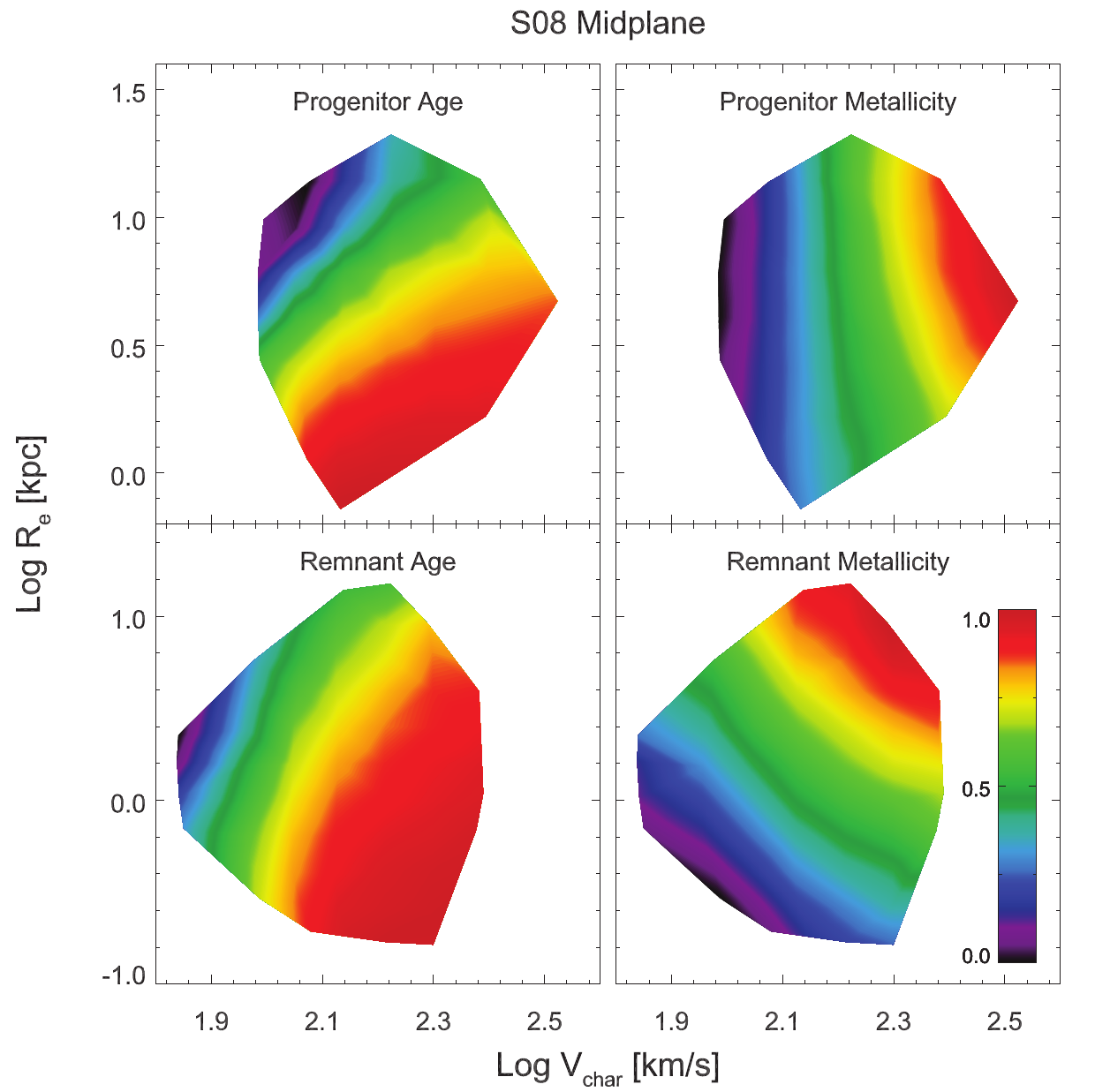}} 
  			}	
		\subfigure{
     			{\includegraphics[width=8cm]{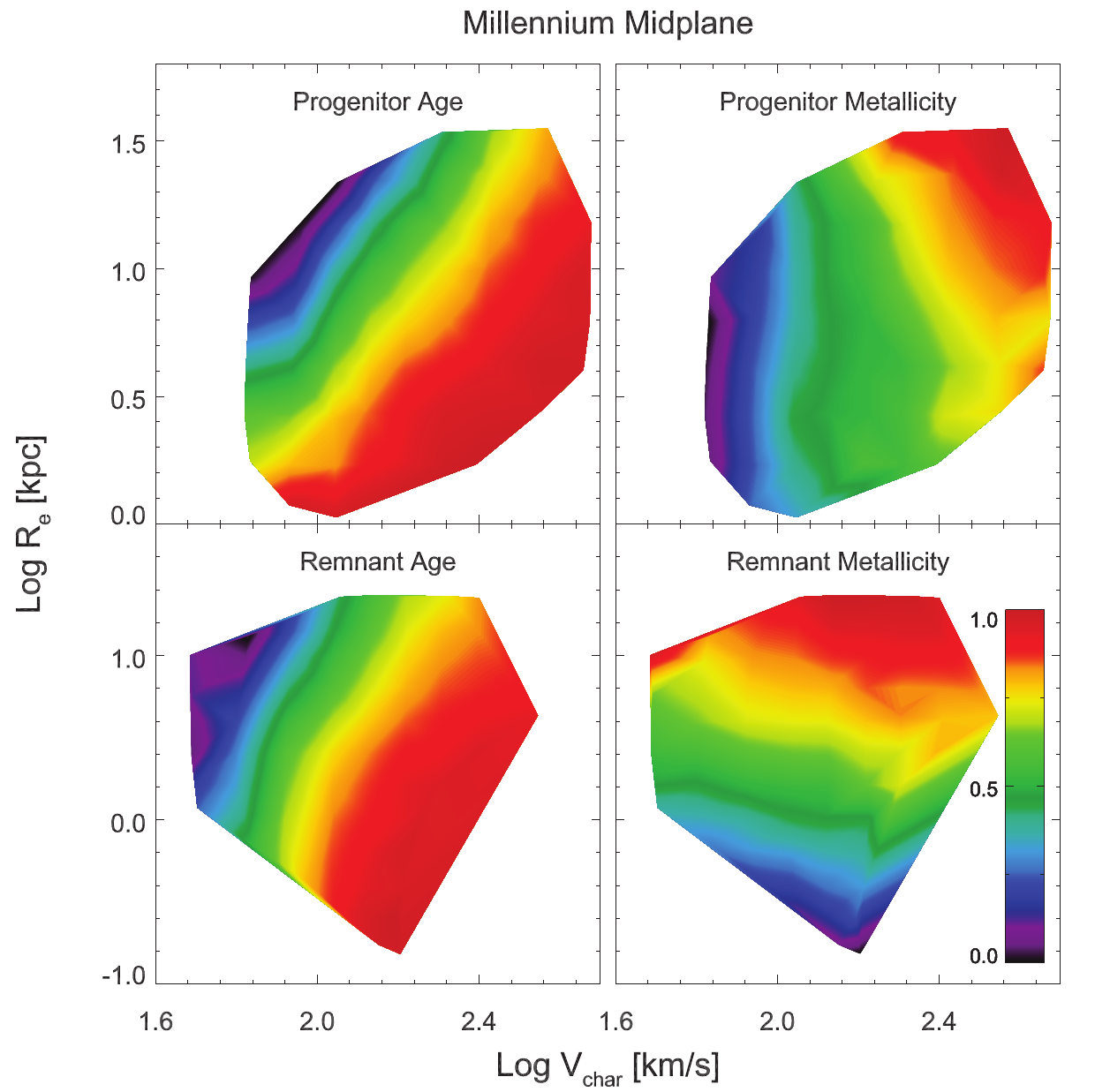}} 
			}
		\caption{Relation between effective radius, characteristic velocity, age (left panels) and metallicity (right panels) for progenitor (top panels) and remnant (bottom panels) galaxies in S08 (top) and Millennium (bottom).  All panels represent the middle slice of the FP and LTFP.  In the progenitor panels, the characteristic velocity is $V_{circ}$, while in the remnant panels the characteristic velocity is $\sigma$.  All of the relations show a counterclockwise rotation between the corresponding relations for the progenitors and remnants.  The contours are individually normalized.}
				\label{fig:fp_angle}
	
\end{figure*}

Comparing these relations with those of the remnants, it is apparent that there is a counterclockwise rotation in the $R_{e}-\sigma$ plane between the comparable relations for progenitors and remnants.  This is made evident in Figure \ref{fig:fp_angle}, in which we plot the age and metallicity FP and LTFP correlations from the midplane alongside each other.  If we form a linear relation between the binned ages and metallicities, $\log R_{e}$, and $\log V_{char}$, where $V_{char} = V_{circ}$ for progenitors and $\sigma$ for remnants, we find that the amount of rotation is $\simeq$ 20\degree counterclockwise for age and metallicity, respectively, in S08.  The amount of rotation in the age-FP correlation is slightly lower for Millennium ($\simeq 10\degree$), while the metallicity-FP correlation exhibits a higher amount of rotation ($\simeq 40\degree$).


\begin{figure*}
	\centering
   		\subfigure{
			{\includegraphics[width=\lbox]{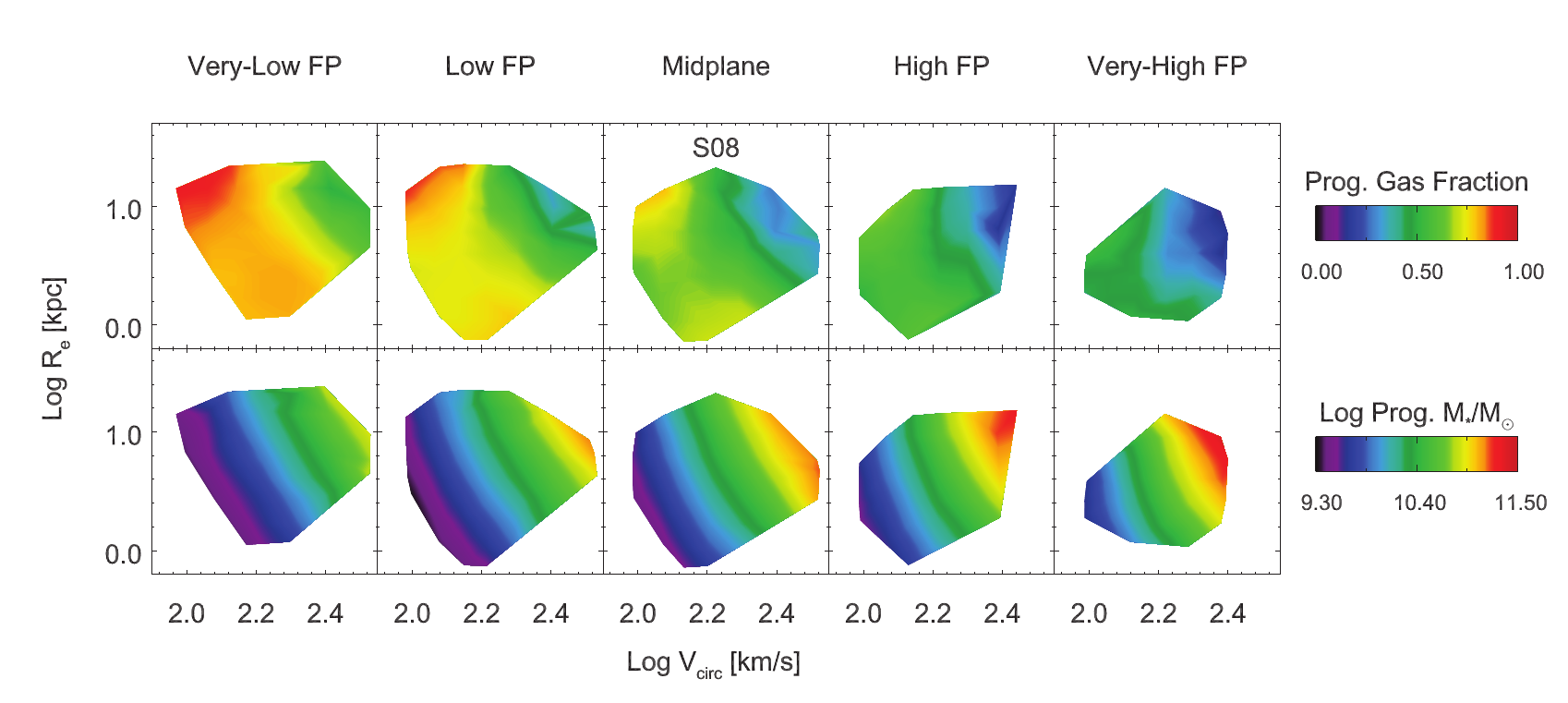}} 
  			}	
		\subfigure{
     			{\includegraphics[width=\lbox]{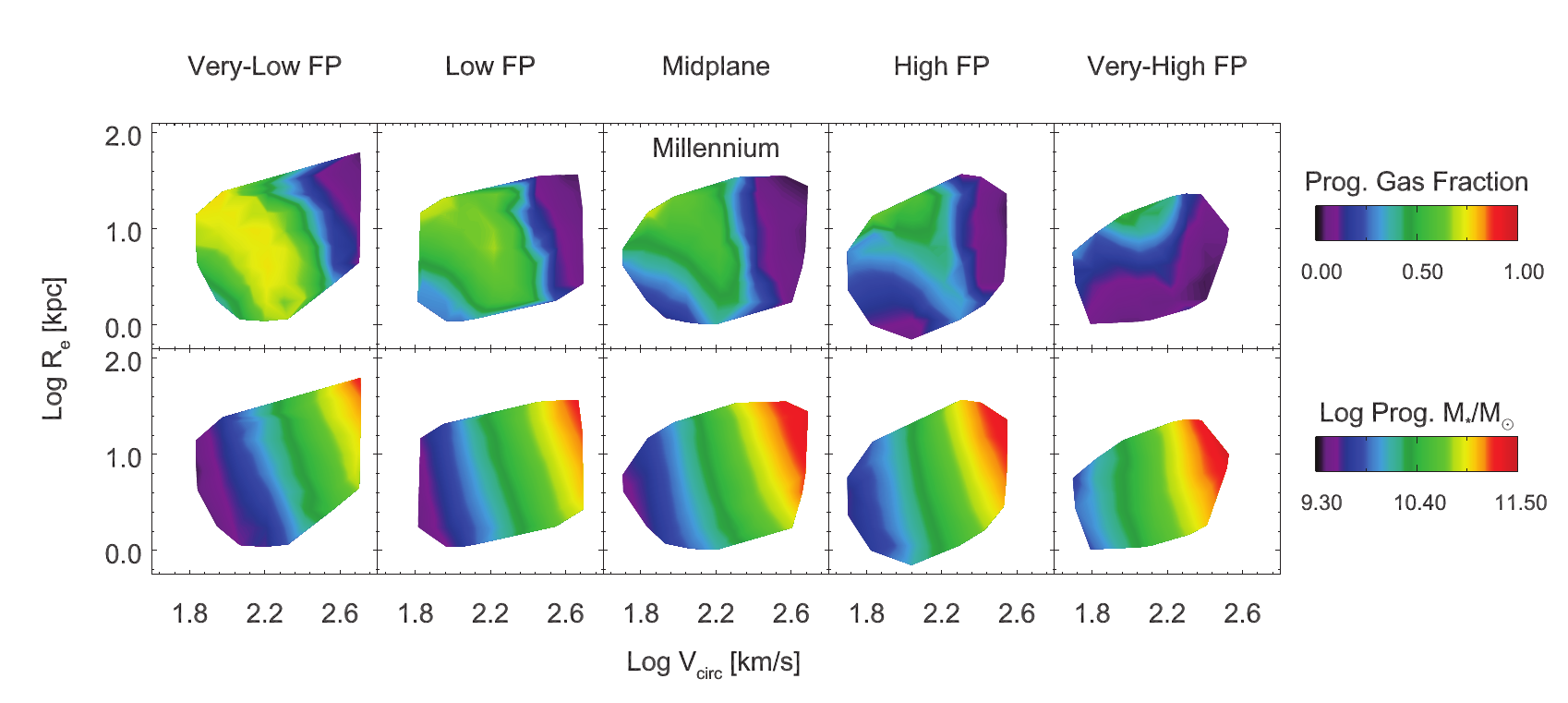}} 
			}
		\caption{Relation between effective radius, circular velocity, gas fraction (top panels) and stellar mass (bottom panels) for spiral merger progenitor galaxies in S08 (top) and Millennium (bottom).  The different panels represent slices of the LTFP.  In both SAMs, the gas fraction is strongly correlated with the progenitor stellar mass; regions with high masses have low gas fractions, and vice versa.  At fixed mass, galaxies with larger radii have lower surface densities and higher gas fractions.}  
				\label{fig:fgas_contour}
	
\end{figure*}

This net rotation is due to the varying gas fractions of the progenitors.  A galaxy's gas contents can be traced to its stellar mass, such that more massive galaxies tend to have lower gas fractions \citep[Figure 10, see also][]{Kannappan:2004a,Catinella:2010a}; furthermore, at fixed mass, the Kennicutt relation implies that galaxies with larger radii have lower surface densities, less efficient star formation and correspondingly higher gas fractions \citep{Covington:2011a}.  Both of these trends are evident when we use the progenitors to produce gas fraction-LTFP and stellar mass-LTFP relations (Figure \ref{fig:fgas_contour}).  The gas fraction is strongly correlated with the stellar mass, with the gas-surface density relation entering as a higher order effect.  

In the dissipational model, galaxies with higher gas fractions undergo more dissipation during the merger, producing remnants that are more compact and have lower velocity dispersions (see Section \ref{ssec:Covmodel}).  Combining this with the dependence of gas fraction on stellar mass and surface density, we find that progenitors with low masses and low surface densities, which lie in the upper left of each of the LTFP slices, tend to contract the most, moving to smaller radii and larger velocity dispersions after the merger.  Conversely, galaxies with high masses and high surface densities have the lowest gas fractions, producing remnants with larger radii and smaller velocity dispersions.  Thus the rotation between the progenitor and remnant properties is largely an effect of gas fraction.  Since the Millennium galaxies have lower gas fractions than the S08 galaxies, the amount of rotation Previous studies have shown that a mass-dependent gas fraction can introduce the observed rotation between the virial plane and the fundamental plane \citep{Dekel:2006a,Robertson:2006a,Hopkins:2008a,Covington:2011a}; here we point out the effects of that rotation on stellar population parameters.

It is worth noting that the amount of star formation has a minimal impact on the ages and metallicities of the remnants; there is a close correspondence between the age and metallicity of the progenitor and remnant, with Spearman rank coefficients greater than $\rho =0.95$ for both parameters and both SAMs.  Thus, the merger has a larger impact on the structural properties of a galaxy than on its stellar population parameters.
 
 \subsection{Comparison to observations}
\begin{figure*}
	\centering
			{\includegraphics[width=10cm]{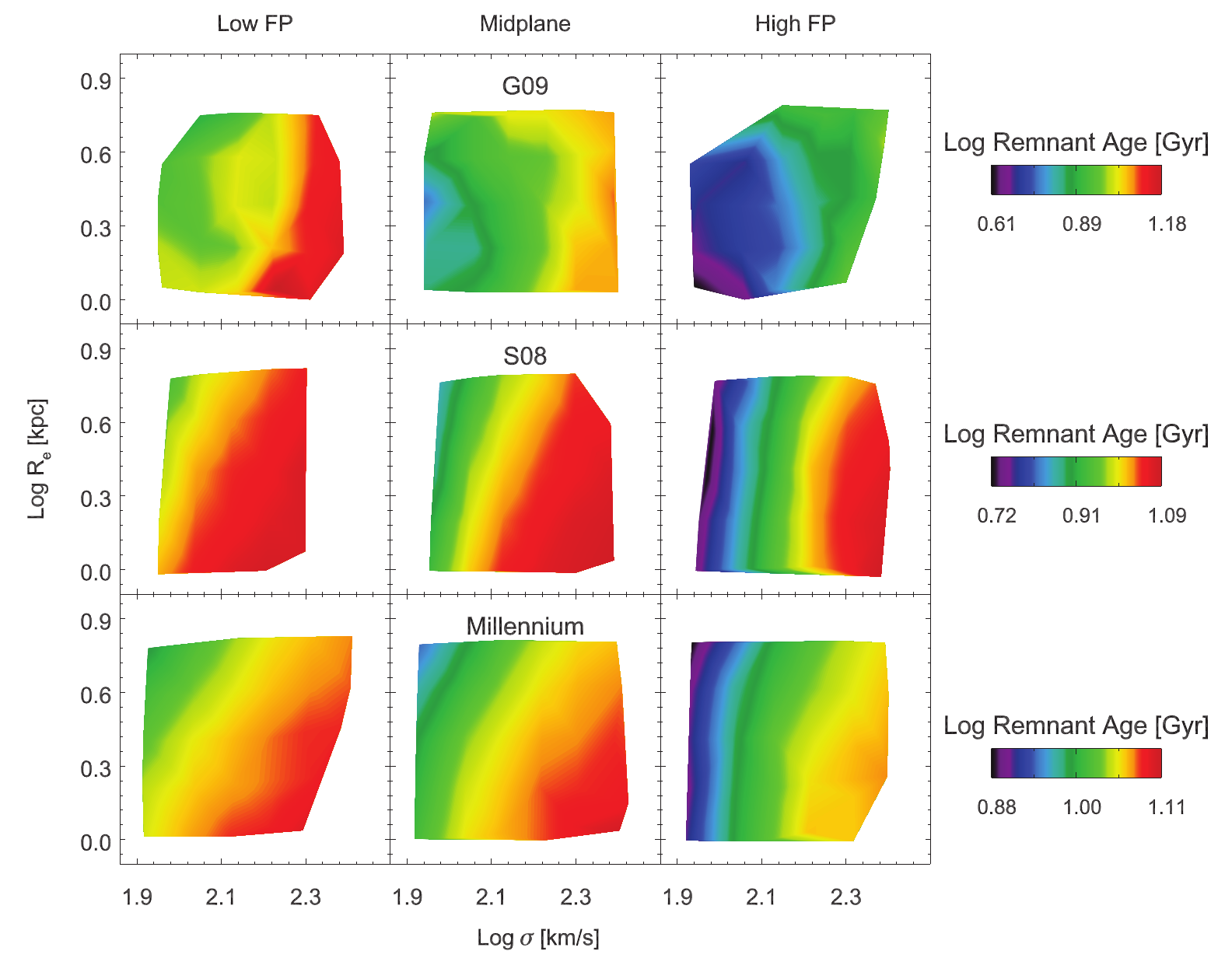}} 	
		\caption{Relation between luminosity-weighted age, effective radius, and velocity dispersion for elliptical galaxies in G09 (top) S08 (middle) and Millennium (bottom).  Here we plot only the region considered in G09.  The different panels represent the three central slices of the FP, as shown in Figure \protect\ref{fig:FP_bin}.  In the SAMs and the observations, stellar population age increases with velocity dispersion, but the simulated galaxies have a further dependence on radius.  Galaxies that lie above the FP also tend to be younger than those that lie below the FP.} 
					\label{fig:age_mid_contour}
	
\end{figure*}

\begin{figure*}
	\centering
		{\includegraphics[width=10cm]{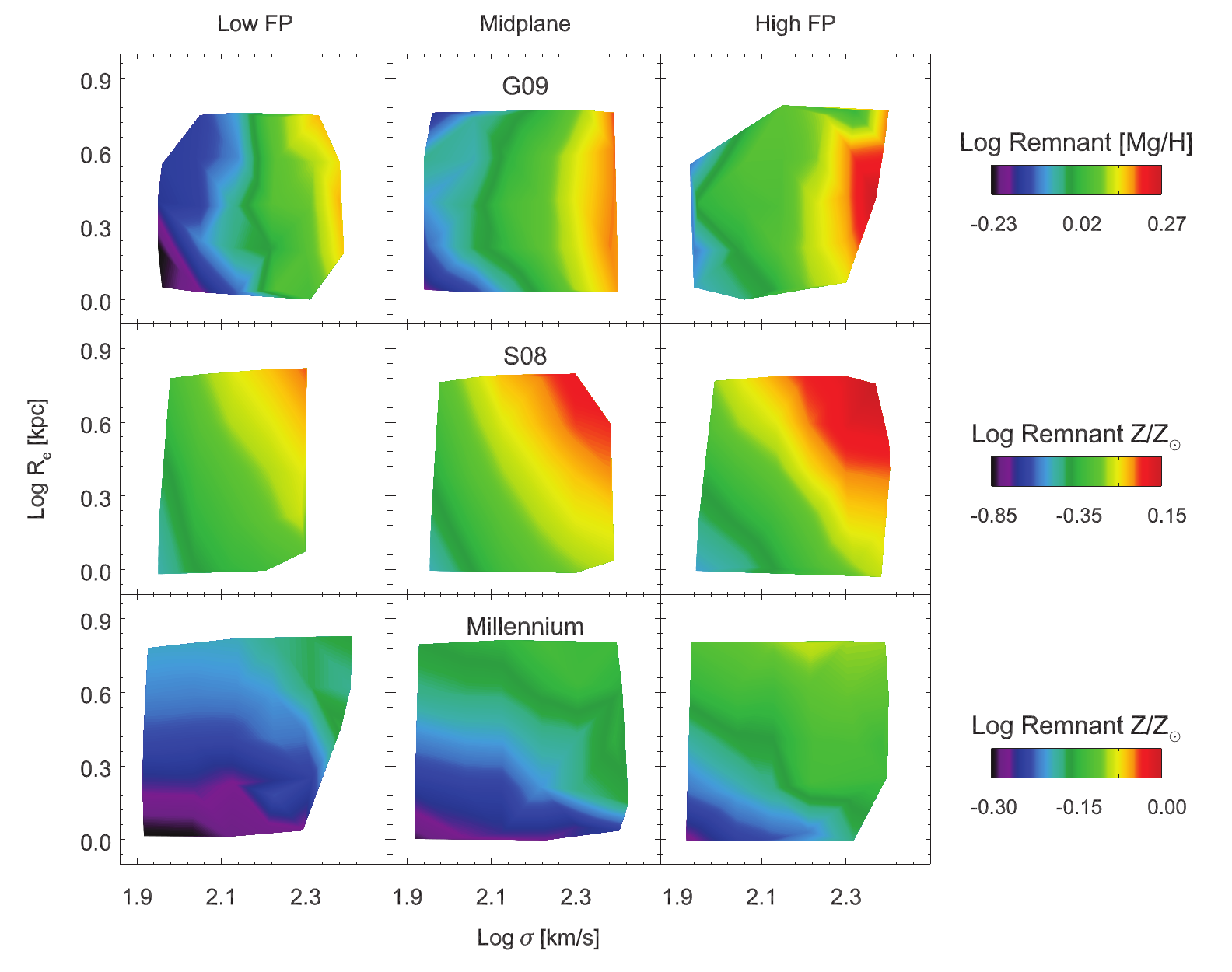}} 
  			
     			\caption{Relation between luminosity-weighted metallicity, effective radius, and velocity dispersion for elliptical galaxies in S08 (top) and Millennium (bottom).  Here we plot only the region considered in G09.  The different panels represent the three central slices of the FP, as shown in Figure \protect\ref{fig:FP_bin}.  While [Mg/H] depends strongly on velocity dispersion in G09, in the SAMs metallicity depends on both velocity dispersion and effective radius.  The SAMs tend to have lower metallicities than observations on average; furthermore, the Millennium SAM produces galaxies with a very narrow range in metallicities.} 
				\label{fig:me_mid_contour}
	
\end{figure*}

 To better compare with G09, we have replotted the age and metallicity contours over the range ($\mathrm{1.9 \ km\ s^{-1} < log \ \sigma
  <2.4\ km \ s^{-1})}$, $\mathrm{(0.0 \ kpc < log\ \emph{R}_{e}   <0.7\ kpc)}$ considered by \cite{Graves:2009b} alongside the G09 data (Figures \ref{fig:age_mid_contour}, and \ref{fig:me_mid_contour}).  We caution that the G09 ages were later found to be systematically high by $\sim 0.12$ dex, owing to weak emission in the H$\beta$ absorption line \citep{Graves:2010a}; however, this would not effect the overall trends.  G09 also calculated luminosity-weighted ages and metallicites using the Lick indices, which have been shown to more closely correlate with the period of most recent star formation, instead of a global quantity \citep{Trager:2009a}.  In future work, we will account for this discrepancy by measuring the Lick indices directly.
  
Examining the trends within FP slices, the age-FP correlations are in rough agreement with G09.  We caution that this agreement occurs in spite of the fact that the progenitors fail to reproduce the observed mass-age relation.  A stronger correlation between age and mass in the progenitors would directly lead to a stronger age-$V_{circ}$ dependence in the LTFP, producing remnants with age-FP correlations that are dependent on both velocity dispersion and radius.  Thus, improving the progenitors' fit to the mass-age relation of \cite{Gallazzi:2005a} would have the effect of worsening the age-FP correlation when compared to G09.  The major difference between our results and those of G09 is that we find metallicity to be dependent on radius and velocity dispersion, while G09 found metallicity to be dependent on velocity dispersion alone. 

Both of the SAMs do a better job of reproducing observed trends through, as opposed to across, the FP.  Galaxies that fall above the FP tend to be younger and more metal-enhanced than average, while those that fall below the FP are older and more metal-poor, in agreement with G09.  This is partially an artifact of truncating star formation after the major merger: galaxies above the FP have universally low merger redshifts while those within and the below the FP have a wide range of merger redshifts.  This results in their having a more recent epoch of star formation and correspondingly younger stellar ages, as well as more extended star formation histories, leading to higher metallicities.  As discussed above however, even at fixed redshift, galaxies above the FP tend to have younger ages and higher metallicities, suggesting that the correlations through the FP are real and not merely artifacts of the methodology.

\section{Evolution with Redshift}
\begin{figure*}
	\centering
		   \subfigure{
		{\includegraphics[width=8.6cm]{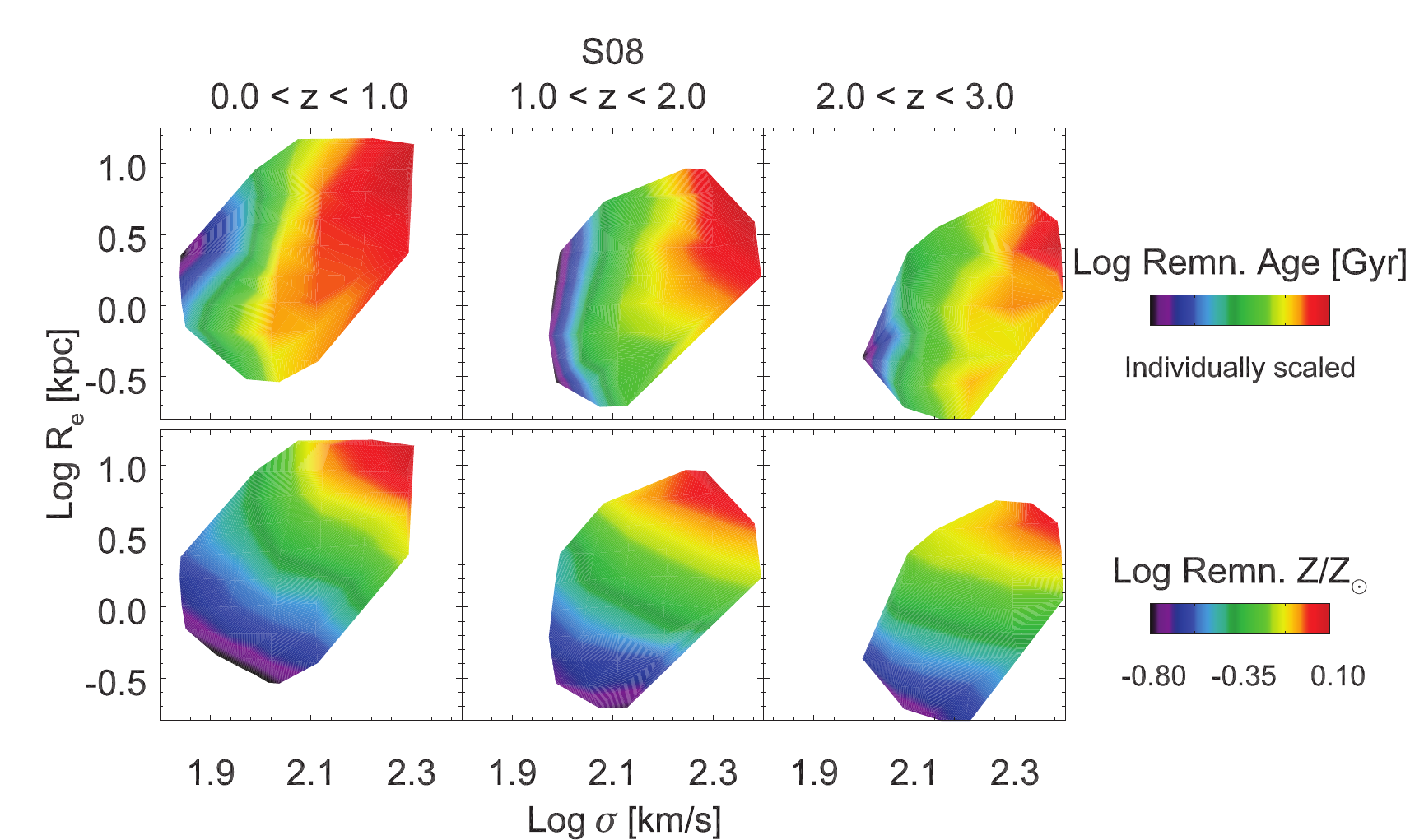}} 
}
	   \subfigure{
		{\includegraphics[width=8.6cm]{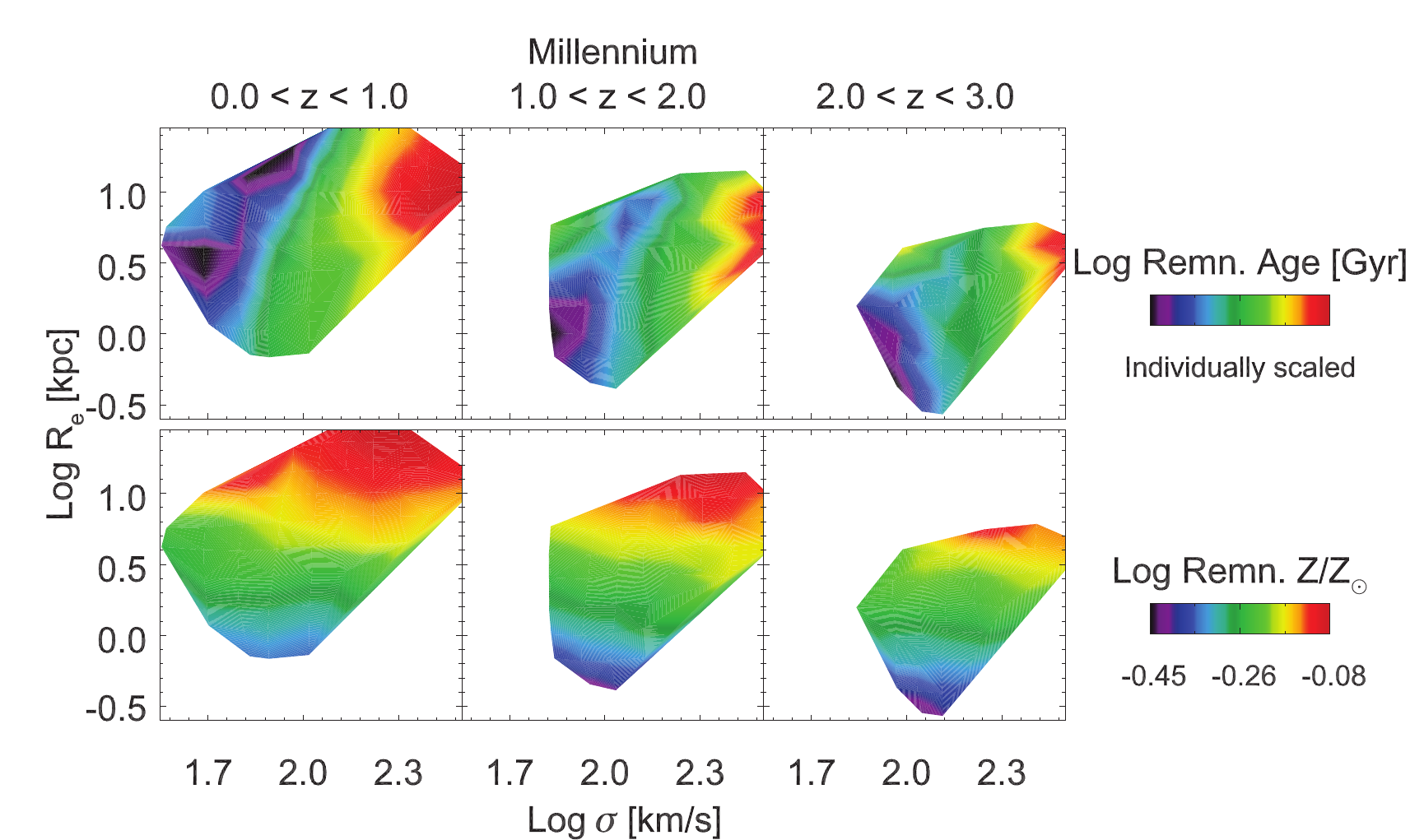}} 
}

		\caption{Relation between luminosity-weighted age (top) and metallicity (bottom), effective radius, and velocity dispersion for S08 (left) and Millennium (right), binned by the redshift of the major merger.  Each panel represents the mid-FP, as shown in Figure \protect\ref{fig:FP_bin}.  Horizontally, panels represent three formation epochs ($ 0.0<z<1.0,\ 1.0<z<2.0,$ and $ 2.0<z<3.0$); each galaxy is only plotted in the redshift bin corresponding to its major merger.  The ages of each redshift panel are individually scaled; from left, they span log $[0.61,1.00], [0.95,1.05], \mathrm{and} [1.03,1.07]$ Gyr for S08 and log $[0.80,1.05], [0.99,1.06], \mathrm{and} [1.05,1.08]$ Gyr for Millennium.  Galaxies that merge at lower redshifts tend to be larger, with lower velocity dispersions.  With the exception of the youngest galaxies within a redshift bin, neither the age-FP relation nor the metallicity-FP relation show an evolution with redshift.}
					 	\label{fig:contour_redshift}
	
\end{figure*}

	The assumption that elliptical remnants experience no further growth or other structural changes following their formation via a major merger has direct implications for the results.  High-redshift observations point to a population of massive, compact ellipticals that are not seen in the local universe \citep{Bell:2004a,Trujillo:2006a,Toft:2007a,Dokkum:2008a,Damjanov:2009a,Franx:2008b,van-Dokkum:2010a,Williams:2010d}.  There is growing evidence that subsequent minor dry mergers may account for at least some of this size evolution, particularly for the most massive galaxies \citep{Naab:2009a,Oser:2010a,Shankar:2010a,Newman:2011a,Shankar:2011a,Trujillo:2011a}, but see also \cite{Nipoti:2009a,Nair:2011a,Saracco:2011a}.  Thus, we expect that the elliptical galaxies with high formation redshifts would undergo significant size evolution by \emph{z}=0.  The contours presented so far stack all ellipticals in a slice, regardless of their formation timescale; here, we divide the population according to the redshift of the major merger when the ellipticals formed.  This allows us to determine whether the redshift of the major merger has an effect on the FP correlations.  Figure \ref{fig:contour_redshift} shows the age and metallicity FP correlations split into 3 redshift bins ($ 0.0<z<1.0,\ 1.0<z<2.0,$ and $ 2.0<z<3.0$).
	
\begin{figure*}
	\centering
		   \subfigure{
		{\includegraphics[width=12cm]{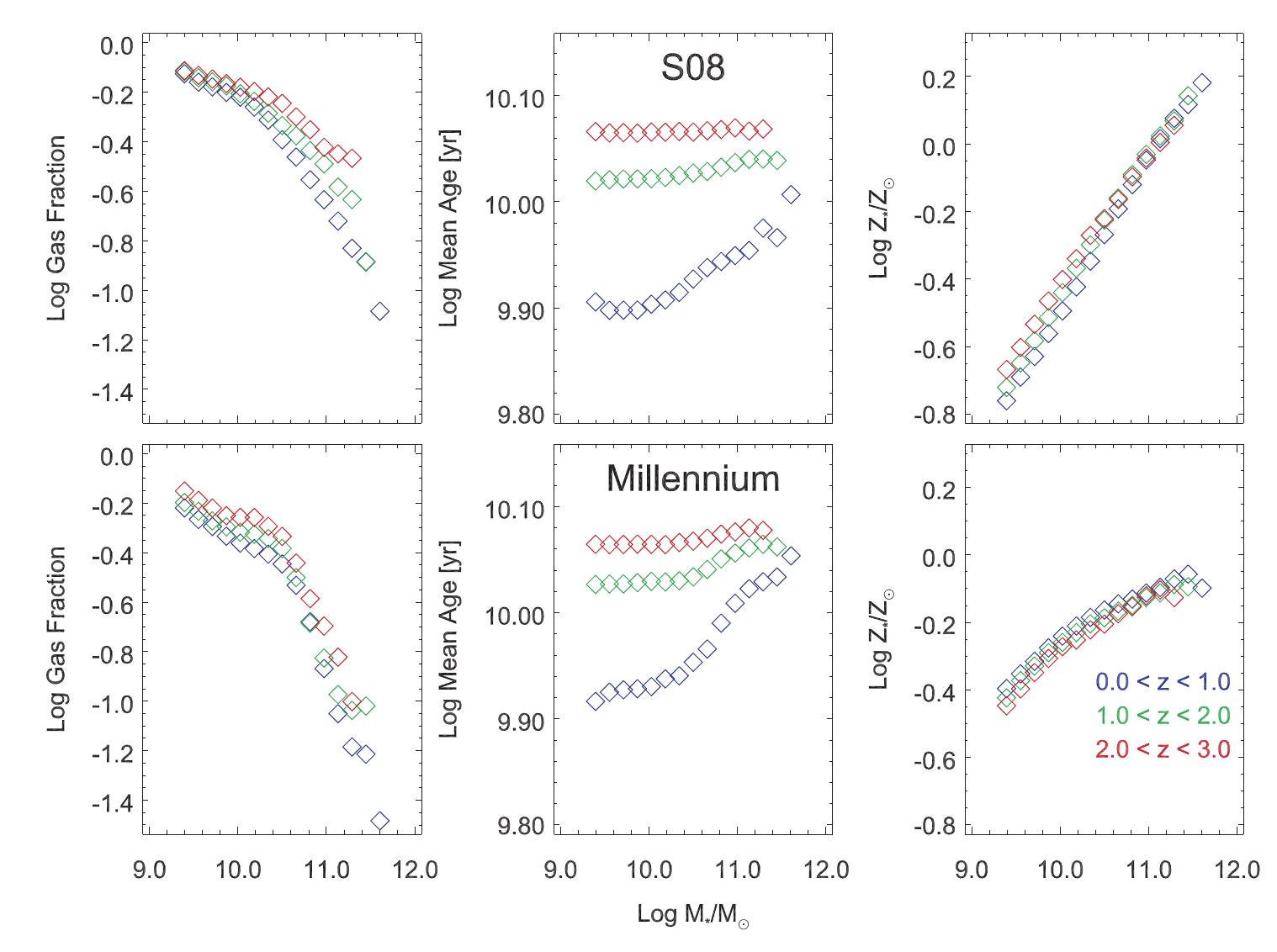}} 
}
		\caption{From left: Evolution of the progenitor stellar gas fraction-, age-, and metallicity-mass correlations for galaxies that merge between z = 3.0 and z = 0.0.  S08 progenitors are shown in the top panels, while Millennium remnants are shown in the bottom panels.  Although both SAMs show offsets such that galaxies that merge at higher redshifts have higher gas fractions and older ages at redshift 0.0, the overall slopes of the relations do not vary with redshift.}
			   	\label{fig:Prog_mass_corrs_z}
	  	
\end{figure*}

	Since galaxies that merge at low redshifts tend to be larger and more gas-poor, the low-redshift elliptical galaxies have both larger radii and smaller velocity dispersions on average.  We find only minimal change in the age-FP trend, and no change in the metallicity-FP trend as a function of redshift.  This is because the dominant factors in producing these trends, namely the relations between stellar mass and age, metallicity, and gas fraction, are similar at all redshifts (Figure \ref{fig:Prog_mass_corrs_z}).  Galaxies that merge at lower redshifts tend to have younger ages and lower gas fractions, but the slopes of the mass-gas fraction and mass-age relations are relatively unchanged.

\subsection{Effects of subsequent evolution}

While the S08 and Millennium SAMs reproduce the observed elliptical mass function \citep{Somerville:2008a,Croton:2006a}, the assumption that galaxies passively evolve following the major merger overproduces low-mass ellipticals and underproduces the high end of the mass function.  This is an expected artifact of the dissipational model: by prohibiting subsequent evolution, we preserve galaxies as they were when they first formed, preventing mass growth through star formation or mergers.  Furthermore, we do not allow for bulge formation through disk instabilities.  Overall, C11 found that applying the dissipational model to major mergers of disk-dominated galaxies alone reproduces only a third of the ellipticals seen in the local universe.  Implementing the model directly within SAMs will reproduce the observed abundances, but this will require the model to be expanded to predict the velocity dispersion and radius after minor and mixed-morphology mergers.

Although we have not attempted to model the evolution of ellipticals, we can estimate its effect.  Dissipationless simulations of major mergers of ellipticals suggest that, while the size-mass relation is preserved, there is increased scatter in radius due to variations in the orbits of the progenitors \citep{Barnes:1992a,Hernquist:1993c,Cole:2000a,Shen:2003a,Boylan-Kolchin:2005a}.  This is generally consistent with C11, who found a size-mass dispersion that was lower than observations.   \cite{Robertson:2006a,Boylan-Kolchin:2006b} demonstrated that major mergers of ellipticals do little to change the tilt of the FP; the increase in the effective radius is offset by similar adjustments in velocity dispersion and surface mass density.  However, recent SPH simulations of dry, minor mergers between a massive elliptical and disk galaxies have suggested that subsequent mergers will greatly increase the radius while slightly decreasing velocity dispersion \citep{Naab:2009a,Oser:2010a,Oser:2012a}. Thus, we suspect that subsequent mergers will weaken the correlations we have found between stellar population parameters and radius, without affecting the trends with velocity dispersion.  This would bring our results into closer agreement with the analysis of G09, although the magnitude of the changes remains unclear.

Alongside the binary major-merger scenario, additional pathways to elliptical galaxy formation have also been suggested.  \cite{Naab:2007a} have proposed minor mergers of disk galaxies, while \cite{Dekel:2009b} point to the importance of ``cold flows'' of gas into disk galaxies.  However, the current generation of SAMs is not equipped to simulate such events \citep{Hirschmann:2011a}.  We hope that including multiple minor mergers and cold flows in the future will improve the accuracy of our predictions.  Both the Millennium and S08 SAMs also allow for bulge formation through disk instabilities, which we have also not attempted to model here.  

\section{Conclusions}
We have used initial conditions for disk-like merger progenitors
predicted by the S08 \citep{Somerville:2008a} and Millennium \citep{Croton:2006a} SAMs along with an analytic model
for the size and velocity dispersion of merger remnants from Covington et al. (2008, 2011) to predict the distribution of these remnants across and through the
fundamental plane. We passively evolve the stellar populations and structural properties of
these remnants to redshift zero, at which point we calculate their
luminosity-weighted ages and metallicites and separate them according
to their residual surface brightness in the fundamental plane.

For both S08 and Millennium, we find that elliptical ages increase as a strong function of velocity dispersion and decrease with radius. The dependence on velocity dispersion is in agreement with G09 while the dependence on radius is not found in observations.  The metallicities, however, increase with both radius and velocity dispersion, whereas the G09 metallicities are independent of radius.  In all cases, the relations for ellipticals are rotated from the corresponding relations for the progenitors.  This rotation arises because the amount of dissipation is strongly dependent on the gas fraction of the progenitor galaxies, which is correlated with the progenitors' masses and circular velocities.  

The difference between the age-FP and metallicity-FP correlations in the remnants stem from the different age-mass and metallicity-mass relations in the progenitors.  While the metallicity-FP correlations are closer to observations in the local Universe than the age-FP correlations, the spiral progenitors are uniformly too old, suggesting that the metallicity-FP correlation is closer to the `true' correlation following a major merger.  The differences between the results using simulated galaxies and the analysis of G09 suggest that a single major merger is not sufficient to remove the correlation between the progenitors' and remnants' radii.

In agreement with G09, we find that galaxies with higher residual surface brightness tend to be younger and more metal-rich.  Examining their structural properties, we find them to have lower stellar mass-to-light ratios and lower central dark matter fractions.  These effects persist even when variations in mass-to-light ratio as a function of merger redshift are considered, suggesting that this is not merely a theoretical artifact.  The galaxies with extremely high surface brightnesses do have lower formation redshifts, suggesting that they will fall to lower FP slices in the future, as their mass-to-light ratios increase.  This is in accordance with \cite{Forbes:1998a,Terlevich:2002a}, and G09.

We find that variations in the stellar mass-to-light ratio and the dark matter fraction within one effective radius contribute to the thickness of the FP, with the dark matter fraction having the larger effect.  If we look across the FP, we find that the two quantities rotate the plane around different axes, with the mass-to-light ratio being highly dependent on velocity dispersion while the dark matter fraction is largely dependent on effective radius.

The present framework is limited in that it assumes that elliptical galaxies are only formed via a single major merger between disk-dominated progenitors, and does not allow for any subsequent growth or mergers. As such, we only produce $\sim 1/3$ of the ellipticals seen in the local universe, with the deficiency being strongest at the high-mass end of the mass function.  In future work, we will incorporate a broader range of processes, including subsequent minor mergers, elliptical-elliptical mergers, and mixed morphology mergers.  We will also separate the red sequence population considered in G09 into elliptical and S0/Sa galaxies, using an automated morphological classification scheme \citep{Cheng:2011a}.

A new version of the Somerville SAM also contains a detailed Galactic Chemical Evolution model, including non-instantaneous recycling, enrichment by both core collapse and Type Ia supernovae, and tracking of multiple chemical elements, as described in \cite{Arrigoni:2010a,Arrigoni:2010b}.  We also plan to calculate ages and metallicities using the Lick indices instead of using approximations to calculate the luminosity-weighted properties.  This will allow us to make direct comparisons with [Fe/H], [Mg/H], and [Fe/Mg].  These improvements will allow us to make more detailed comparisons to observations.
\section{Acknowledgments}
We thank Brad Holden, Patrik Jonsson, Mark Krumholz, Thorsten Naab, Stefano Profumo, and Connie Rockosi for several useful discussions.  LAP thanks the Space Telescope Science Institute for support and hospitality. DC acknowledges receipt of a QEII Fellowship from the Australian Government.  Research of LAP and JRP was supported by NSF AST-1010033 and HST-GO-12060.12-A. 
\bibliographystyle{monthly}
\bibliography{bib}

\end{document}